\documentclass[prd,rd,showpacs,twocolumn,superscriptaddress]{revtex4-1}
\usepackage{amsfonts,amsmath,amssymb,amsthm}
\usepackage[bookmarks, bookmarksopen, bookmarksnumbered]{hyperref}

\usepackage{graphicx}
\usepackage{color}
\usepackage{siunitx}
\usepackage[utf8]{inputenc}
\usepackage[T1]{fontenc}
\usepackage{lmodern}
\usepackage[right]{rotating}
\usepackage{array}
\usepackage{multirow}
\usepackage{paralist}
\usepackage{colortbl}

\usepackage{hyperref}
\hypersetup{
    colorlinks=true,
    linkcolor=blue,
    filecolor=magenta,      
    urlcolor=magenta,
    citecolor=blue,
    pdfpagemode=FullScreen,
    }

\usepackage{physics}
\usepackage{wasysym}
\usepackage{tensor}
\usepackage{multirow}
\numberwithin{equation}{section}
\usepackage{enumitem}
\setlist{leftmargin=*}
\usepackage{siunitx}
\newcommand{\parcite}[1]{\cite{#1}}

\newcommand{\MoL}{\texttt{MoL} }
\newcommand{\MIR}{\texttt{MIR}}
\newcommand{\en}{\mathcal{E}}
\newcommand{\tensore}[1]{\vb*{\overline{#1}}}
\newcommand{\BB}{\vb{B}}
\newcommand{\EE}{\vb{E}}
\newcommand{\JJ}{\vb{J}}
\newcommand{\Mcon}{\vb{M}}
\newcommand{\Scon}{\vb{S}}
\newcommand{\vv}{\vb{v}}

\newcommand{\Uem}{U_{em}}
\newcommand{\XX}{\vb*{\mathcal{X}}}
\newcommand{\YY}{\vb*{\mathcal{Y}}}
\newcommand{\QQ}{\vb*{\mathcal{Q}}}
\newcommand{\RR}{\vb*{\mathcal{R}}}
\newcommand{\UU}{\vb*{\mathcal{U}}}
\newcommand{\UUbar}{\vb*{\overline{{\mathcal{U}}}}}
\newcommand{\LL}{\vb*{\mathcal{L}}}
\newcommand{\zerovec}{\vb{0}}
\newcommand{\at}{\tilde{a}}
\newcommand{\bt}{\tilde{b}}
\newcommand{\ut}{\tilde{u}}
\newcommand{\etat}{\tilde{\eta}}
\newcommand{\FF}{\mathcal{F}}
\newcommand{\PP}{\vb{P}}

\newcommand{\bvec}{\vb{b}}
\newcommand{\evec}{\vb{e}}
\newcommand{\rvec}{\vb{r}}

\newcommand{\qbar}{\bar{q}}

\usepackage[normalem]{ulem}

\begin{document}

%\title{On the properties of the post-merger gravitational wave signal emitted by binary neutron star 
%mergers when the remnant is long living}
\title{MIR: a general-relativistic resistive-magneto-hydrodynamic code to study the effect of resistivity in Neutron Star dynamics}
\date{\today}

%%% -----------------------------

\author{Kevin \surname{Franceschetti}}
\affiliation{Parma University, Parco Area delle Scienze 7/A, I-43124 Parma (PR), Italy}
\affiliation{INFN gruppo collegato di Parma, Parco Area delle Scienze 7/A, I-43124 Parma (PR), Italy}

\author{Roberto \surname{De Pietri}}
\affiliation{Parma University, Parco Area delle Scienze 7/A, I-43124 Parma (PR), Italy}
\affiliation{INFN gruppo collegato di Parma, Parco Area delle Scienze 7/A, I-43124 Parma (PR), Italy}

\begin{abstract}
    Electrical resistivity plays a fundamental role in many astrophysical systems, influencing the evolution of the magnetic field and energy dissipation processes. During the coalescence of two neutron stars, resistive effects can significantly affect the dynamics and gravitational and electromagnetic signatures associated with these events, such as gamma-ray bursts and kilonova emission. Here, we present our new code, named \MIR. Developed within the EinsteinToolkit framework, \MIR{} solves the general relativistic magnetohydrodynamic equations in 3D Cartesian coordinates and on a dynamical spacetime using the 3+1 Eulerian formalism, in both the ideal and resistive regimes, filling a crucial gap in the toolkit's capabilities.
\end{abstract}

% The next command needs to come _after_ the abstract to pick up the
% columnwidth in two-column mode
%\LTcapwidth=\columnwidth


\maketitle

%%%%%%%%%%%%%%%%%%%%%%%%%%%%%%%%%%%%%%%%%%%%%%%%%%%%%%%
\section{Introduction}

General relativistic magnetohydrodynamics (GRMHD) is a powerful tool for studying a wide range of astrophysical phenomena, from the formation of black holes (BHs) to the evolution of neutron stars (NSs) and binary neutron star (BNS) systems. In recent years, most GRMHD simulations - in particular of BNS systems - have been carried out in the so-called ideal magnetohydrodynamic (IMHD) regime, in which the electrical resistivity - the ability of a material to resist the flow of electric current - is negligible (e.g., \cite{Duez2005,Shibata2005,Rezzolla2011,Endrizzi2016,Baiotti2017,Ciolfi2017,Ciolfi2019,Ciolfi2020,Sur2022}). In this regime, magnetic field lines are conserved, meaning they cannot be created or destroyed. Even if the IMHD is a good approximation for many astrophysical phenomena, in some cases, electrical resistivity is not negligible, and IMHD is no longer valid. In these cases, it is necessary to use resistive MHD (RMHD), which includes electrical resistivity in the equations.

In astrophysical plasmas, resistivity can be caused by various factors, including collisions between particles, impurities, and turbulence. The inclusion of resistivity in GRMHD simulations is important for a number of reasons. First, it can lead to the formation of new phenomena, such as magnetic reconnection \parcite{Perez-Coll2022}. Second, it can affect the evolution of existing phenomena, such as the formation of jets and accretion disks. Third, it can play a role in the dynamics of NSs and BHs. In an NS merger, the merger of the two magnetic fields can generate a large amount of magnetic energy, which can be dissipated by electrical resistivity. This dissipation can lead to the formation, stabilization, and collimation of relativistic jets, which are responsible for the gamma-ray emission. Electrical resistivity can also convert magnetic energy into thermal energy, which can lead to the emission of electromagnetic radiation at longer wavelengths, such as kilonova emission.

To date, general relativistic RMHD simulations have been performed to study the evolution of accretion disks \parcite{Bugli2014,Tomei2019}, isolated neutron stars \parcite{DelZanna2022, Franceschetti2020,Cheong2025}, relativistic jets \parcite{Mattia2023}, and the dynamics of relativistic reconnection \parcite{Bugli2024}. Noteworthy are the studies of \parcite{Dionysopoulou2015} and \cite{Shibata2021}, which performed resistive simulations of neutron star mergers, the latter incorporating the dynamo term following the approach presented in \cite{Bucciantini2013}. However, the numerical codes used in these works are not publicly available.

In many numerical simulations, the spatial resolution - limited by the available computational resources - is insufficient to capture very small details or high-frequency features of the phenomenon being studied. It would be necessary to use extremely fine grids to resolve all scales, which would result in a prohibitive computational cost. Moreover, MHD equations are hyperbolic and prone to developing discontinuities such as shocks and tangential discontinuities. These phenomena, if not handled with appropriate numerical techniques, can generate spurious oscillations due to the Gibbs phenomenon and numerical instabilities, compromising the accuracy of the results. The lack of physical dissipation in the IMHD equations allows magnetic structures to collapse to ever smaller scales, exceeding the numerical grid resolution and leading to the accumulation of kinetic and magnetic energy. This phenomenon, known as the \textit{energy cascade towards small scales}, inevitably leads to the divergence of the numerical code.
Additionally, numerical discontinuities or shock waves can become very steep and cause unstable behavior, such as spurious oscillations or non-physical solutions. To handle these numerical problems and ensure simulation stability, an artificial viscosity - such as the fifth-order Kreiss-Oliger dissipation \parcite{Kreiss1973} - is added to IMHD simulations. This term, absent in nature, aims to dissipate the 'energy' present or introduced by numerical method on the smallest scales. The choice of the artificial viscosity coefficient represents a delicate compromise: an excessive value excessively smoothes the physical structures, while a too-low value is insufficient to guarantee numerical stability. Including physical dissipation mechanisms, such as physical viscosity or electrical resistivity, should dampen small-scale discontinuities or instabilities, eliminating the need to add artificial dissipation. Furthermore, in nature, there are always some dissipative mechanisms (albeit very small) that act on very small scales. Thus, including physical dissipation not only helps numerically but also adds physical realism to the simulation.

In this paper, we present our new numerical code, \MIR{} (an acronym for MagnetoIdrodinamica Resistiva, i.e., resistive magnetohydrodynamics in Italian), written in Fortran90 and developed to operate within the Einstein Toolkit framework \parcite{Loffler2012}, capable of solving the equations of general relativistic resistive magnetohydrodynamics (RMHD) in a three-dimensional dynamical spacetime.

The paper is organized as follows: in Section \ref{sec2} we present the general relativistic RMHD equations; in Section \ref{sec3} we describe the numerical scheme and its implementation in the EinsteinToolkit \parcite{Loffler2012}; in Section \ref{sec4} we present the results of our tests; finally, in Section \ref{sec5} we summarise and discuss future developments.

In the following, we assume to use Greek letters (running from 0 to 3) for 4D space-time components, Latin letters (running from 1 to 3) for the 3D spatial components, and a signature (-,+,+,+) for the space-time metric. Bold characters are used for vectors (e.g., $\vv$), while bold characters with an overlaid line are used for tensors (e.g., $\tensore{T}$). Moreover, we set $G=c=M_{\astrosun}=1$ (with $M_{\astrosun}$ the solar mass) and the Lorentz-Heaviside units ($\varepsilon_0=\mu_0=1$) are used for the electromagnetic quantities.

%%%%%%%%%%%%%%%%%%%%%%%%%%%%%%%%%%%%%%%%%%%%%%%%%%%%%%%
\section{Governing equations}\label{sec2}
The study of the time evolution of a magnetized fluid requires solving both the Euler and Maxwell equations. While these are well-known and widely referenced in the literature (e.g. \cite{DelZanna2007,Mosta2014,Etienne2015,Cipolletta2020,Giacomazzo2007}), we include them here for completeness, presenting both the covariant approach and the 3+1 formalism \parcite{Alcubierre2008,Baumgarte2010,Gourgoulhon2012}. The latter is particularly useful from a computational perspective, as it separates spatial and temporal variables. However, the number of equations is insufficient to account for all the physical variables involved. Therefore, to close the system, it is necessary to introduce additional relations, which are also presented in this section.

\subsection{Covariant approach}
The evolution of a fluid interacting with an electromagnetic field is described by Euler's equations
\begin{equation}
    \nabla_{\mu}\qty(\rho u^{\mu})=0
    \label{eq:Euler-rho-covariant}
\end{equation}
\begin{equation}
    \nabla_{\mu}T^{\mu\nu}=0
    \label{eq:Euler-Tmunu-covariant}
\end{equation}
and Maxwell equations
\begin{equation}
    \nabla_{\mu}F^{\mu\nu}=-I^{\nu}
    \label{eq:Maxwell-Fmunu-covariant}
\end{equation}
\begin{equation}
    \nabla_{\mu}F^{*\mu\nu}=0
    \label{eq:Maxwell-Fstarmunu-covariant}
\end{equation}
Here $\nabla_{\mu}$ is the space-time covariant derivative, $\rho$ the mass density as measured in the frame comoving with 4-velocity $u^{\mu}$, $T^{\mu\nu}$ the total momentum-energy tensor, $F^{\mu\nu}$ the electromagnetic Faraday tensor, $F^{*\mu\nu}$ its dual, and $I^{\mu}$ the 4-vector current density. The $T^{\mu\nu}$ tensor is given by the sum of two contributions, one due to matter
\begin{equation}
    T^{\mu\nu}_m=\rho hu^{\mu}u^{\nu}+pg^{\mu\nu}
    \label{eq:Tmunu-matter-covariant}
\end{equation}
- where $h=1+\epsilon+p/\rho$ is the specific enthalpy, $\epsilon$ the specific internal energy, $p$ the thermal pressure, and $g_{\mu\nu}$ the metric tensor - and one due to the electromagnetic field
\begin{equation}
    T^{\mu\nu}_{em}=\tensor{F}{^{\mu}_{\lambda}}F^{\nu\lambda}-\dfrac{1}{4}\qty(F^{\lambda\kappa}F_{\lambda\kappa})g^{\mu\nu}
    \label{eq:Tmunu-em-covariant}
\end{equation}
The Faraday tensor, its dual, and the current density can be decomposed as
\begin{equation}
    F^{\mu\nu}=u^{\mu}e^{\nu}-e^{\mu}u^{\nu}+\varepsilon^{\mu\nu\lambda\kappa}b_{\lambda}u_{\kappa}
    \label{eq:Fmunu-covariant}
\end{equation}
\begin{equation}
    F^{*\mu\nu}=u^{\mu}b^{\nu}-b^{\mu}u^{\nu}-\varepsilon^{\mu\nu\lambda\kappa}e_{\lambda}u_{\kappa}
    \label{eq:Fstarmunu-covariant}
\end{equation}
and
\begin{equation}
    I^{\mu}=q_0u^{\mu}+j^{\mu}
    \label{eq:Imu-covariant}
\end{equation}
where $\varepsilon^{\mu\nu\lambda\kappa}$ is the space-time Levi-Civita tensor density, and $e^{\mu}$, $b^{\mu}$, $q_0$, and $j^{\mu}$ are the electric field, magnetic field, charge density, and (conduction) current measured in the frame comoving with 4-velocity $u^{\mu}$.

A possible closure for Maxwell equations is given by the (isotropic) Ohm's law, whose expression in general relativity is
\begin{equation}
    j^{\mu}=\sigma e^{\mu}
    \label{eq:Ohm-sigma}
\end{equation}
where $\sigma$ is the conductivity tensor, related to the microphysics of the plasma via \parcite{Bekenstein1978,Zanotti2011}
\begin{equation}
    \sigma=n_e e^2 \tau_c m_e^{-1}
    \label{eq:sigma-coefficient}
\end{equation}
Here $e$ is the electron's charge, $m_e$ the electron's mass, $\tau_c$ the collision time. Since we do not have access to the microphysics of the plasma, we treat $\sigma$ as a free parameter of the theory. The IMHD regime corresponds to $\sigma\to\infty$ and $e^{\mu}=0$.

\subsection{3+1 decomposition}
As in many numerical codes used to study the time evolution of astrophysical systems in general relativity - such as \texttt{ECHO} \parcite{DelZanna2007}, \texttt{GRHydro} \parcite{Mosta2014}, \texttt{IllinoisGRMHD} \parcite{Etienne2015}, \texttt{Spritz} \parcite{Cipolletta2020}, \texttt{WhiskyMHD} \parcite{Giacomazzo2007}, and \texttt{Gmunu} \parcite{Cheong2020,Cheong2022} - we adopt the 3+1 formalism \parcite{Alcubierre2008,Baumgarte2010,Gourgoulhon2012}. The form of the element line is 
\begin{equation}
    \dd s^2=-\qty(\alpha^2-\beta^i\beta_i) \dd t^2+2\beta_i \dd x^i\dd t+\gamma_{ij}\dd x^i \dd x^j
    \label{eq:ds2}
\end{equation}
where $\alpha$ is called the lapse function, $\beta^i$ is called the shift vector, and $\gamma_{ij}$ is the spatial 3-metric with determinant $\gamma$. Each 4-vector and 4-tensor defined above can be decomposed as
\begin{equation}
    u^{\mu}=Wn^{\mu}+Wv^{\mu}
    \label{umu-3+1}
\end{equation}
\begin{equation}
    T^{\mu\nu}=\en n^{\mu}n^{\nu}+S^{\mu}n^{\nu}+S^{\nu}n^{\mu}+S^{\mu\nu}
    \label{Tmunu-3+1}
\end{equation}
\begin{equation}
    F^{\mu\nu}=n^{\mu}E^{\nu}-E^{\mu}n^{\nu}+\varepsilon^{\mu\nu\lambda\kappa}B_{\lambda}n_{\kappa}
    \label{Fmunu-3+1}
\end{equation}
\begin{equation}
    F^{*\mu\nu}=n^{\mu}B^{\nu}-B^{\mu}n^{\nu}-\varepsilon^{\mu\nu\lambda\kappa}E_{\lambda}n_{\kappa}
    \label{Fstarmunu-3+1}
\end{equation}
\begin{equation}
    I^{\mu}=qn^{\mu}+J^{\mu}
    \label{Imu-3+1}
\end{equation}
where
\begin{equation}
    n_{\mu}=\qty(-\alpha,0_i) \qquad n^{\mu}=\qty(1/\alpha,-\beta^i/\alpha)
\end{equation}
is the time-like unit vector corresponding to the 4-velocity of the so-called \textit{Eulerian observer}, $v^{\mu}$ the usual fluid velocity, $W$ the Lorentz factor, $\en$ the total energy density, $S^{\mu}$ the total momentum, $E^{\mu}$ the electric field, $B^{\mu}$ the magnetic field, $q$ the electric charge density, and $J^{\mu}$ the current density as measured by the Eulerian observer. Euler and Maxwell equations become
\begin{equation}
    \partial_t\qty(\sqrt{\gamma}\, D)+\sqrt{\gamma}\div\qty[D\qty(\alpha\vv-\vb*{\beta})]=0
    \label{eq:mass_conservation}
\end{equation}
\begin{equation}
    \partial_t\qty(\sqrt{\gamma}\, \en)+\sqrt{\gamma}\div\qty(\alpha\Scon-\en\vb*{\beta})=\sqrt{\gamma}\qty(\alpha\tensore{S}\boldsymbol{:}\tensore{K}-\Scon\vdot\grad\alpha)
    \label{eq:energy_conservation}
\end{equation}
\begin{equation}
    \partial_t\qty(\sqrt{\gamma}\, \Scon)+\sqrt{\gamma}\div\qty(\alpha\tensore{S}-\vb*{\beta}\Scon)=\sqrt{\gamma}\qty[\Scon\vdot\qty(\grad\vb*{\beta})-\en\grad\alpha]
    \label{eq:momentum_conservation}
\end{equation}
\begin{equation}
    \partial_t\qty(\sqrt{\gamma}\,\BB)+\sqrt{\gamma}\curl\qty(\alpha\EE+\vb*{\beta}\cp\BB)=\vb{0}
    \label{eq:Maxwell3}
\end{equation}
\begin{equation}
    \partial_t\qty(\sqrt{\gamma}\,\EE)-\sqrt{\gamma}\curl\qty(\alpha\BB-\vb*{\beta}\cp\EE)=\sqrt{\gamma}\qty(-\alpha\JJ+q\vb*{\beta})
    \label{eq:Maxwell4}
\end{equation}
\begin{equation}
    q=\div\EE
    \label{eq:Maxwell1}
\end{equation}
\begin{equation}
    \div\BB=0
    \label{eq:Maxwell2}
\end{equation}
with
\begin{equation}
    D=W\rho
    \label{eq:D-definition}
\end{equation}
\begin{equation}
    \Scon=\rho hW^2\vv+\EE\cp\BB=\Mcon+\EE\cp\BB
    \label{eq:S-definition}
\end{equation}
- where $\Mcon$ is the fluid momentum -
\begin{equation}
    \en=\rho hW^2-p+\qty(E^2+B^2)/2=\en_F+\Uem
    \label{eq:En-definition}
\end{equation}
- where $\en_F$ is the term due to the matter and $\Uem$ the electromagnetic energy density -
\begin{equation}
    \tensore{S}=\rho hW^2\vv\vv-\EE\EE-\BB\BB+\qty(p+\Uem)\tensore{\vb*{\gamma}}
    \label{eq:Sij-definition}
\end{equation}
and $\tensore{K}$ is the \textit{extrinsic curvature}. The 3+1 decomposition of Eq. \eqref{eq:Ohm-sigma} gives
\begin{equation}
    \JJ=q\vv+\dfrac{W\qty[\EE+\vv\cp\BB-\qty(\EE\vdot\vv)\vv]}{\eta}
    \label{eq:J-vector}
\end{equation}
where
\begin{equation}
    \eta=1/\sigma
    \label{eq:eta-definition}
\end{equation}
is the electrical resistivity. In the IMHD regime, the electric field is not evolved but is given by the relation
\begin{equation}
    \EE=-\vv\cp\BB
    \label{eq:E-imhd}
\end{equation}

%%%%%%%%%%%%%%%%%%%%%%%%%%%%%%%%%%%%%%%%%%%%%%%%%%%%%%%
\section{Numerical scheme} \label{sec3}
\subsection{Implementation of the IMEX scheme in the Einstein Toolkit}
The set of equations described in the previous section can be rewritten as
\begin{equation}
    \partial_t\UU=-\partial_k\vb*{\FF}^k+\vb*{\mathcal{S}}=\QQ+\RR
    \label{eq:conservatives_equations}
\end{equation}
where
\begin{equation}
    \partial_t\UU=\sqrt{\gamma}\qty[D,S_i,\en,B^i,E^i]^T
    \label{eq:conserved_variables}
\end{equation}
are the \textit{conserved} variables, $\vb*{\FF}$ the corresponding fluxes, $\vb*{\mathcal{S}}$ the corresponding source terms, $\QQ$ the right-hand side (RHS) without stiff terms (including flux derivatives) and $\RR$ the RHS containing the stiff terms, i.e. those proportional to $\eta^{-1}\gg 1$. If $\RR=\zerovec$, Eq. \eqref{eq:conservatives_equations} reduces to the usual ideal equations and the usual Runge-Kutta (RK) schemes can be used to solve them. However, if $\RR\neq\zerovec$ an IMEX (\textit{IMplicit-EXplicit}) scheme is needed \parcite{Pareschi2005,Palenzuela2009}. As reported in Appendix \ref{app_IMEX}, the IMEX scheme requires different coefficients for the two terms on the RHS. This means that, in general, we have to register two RHSs instead of one. However, the time integration in the Einstein Toolkit is handled by the \MoL package (or "thorn"), which requires only one RHS to be registered. In this section, we present a way to implement the SSP2(2,2,2) scheme, which is the IMEX extension of the second-order Runge-Kutta (RK2) scheme, by registering only one RHS. The Butcher tableau \parcite{Butcher1987} for this scheme is reported in Table \ref{tab1}. Other IMEX schemes (up to the third order) are presented in \cite{Pareschi2005} and \cite{Palenzuela2009}.
\begin{table}[t]
    \centering
    \begin{tabular}{c|cccc|cccc}
         0&  0&  0&  &  $\gamma$&  $\gamma$& 0& &\multirow{3}{*}{$\gamma=1-\dfrac{1}{\sqrt{2}}$}\\ 
         1&  1&  0&  &  $1-\gamma$&  $1-2\gamma$& $\gamma$& &\\ \cline{1-3}\cline{5-7}
         &  1/2&  1/2&  &  &  1/2& 1/2& &\\
    \end{tabular}
    \caption{Tableau for the explicit (left) implicit (right) IMEX-SSP2(2,2,2) L-stable scheme \parcite{Pareschi2005,Palenzuela2009}.}
    \label{tab1}
\end{table}

We denote by $\UU_*^{(i)}$ the value of the conserved variable $\UU$ returned by the \MoL thorn after the $i$-th explicit step, by $\UUbar^{(i)}$ the scratch array defined in the $i$-th step, and by $\LL_i=\QQ_i+\RR_i$ the total RHS evaluated after the $i$-th step, i.e., $\LL_i=\LL\qty[\UU^{(i)}]$ (the same for $\QQ_i$ and $\RR_i$).

For the SSP2(2,2,2) scheme, the following three steps are required (one more than the RK2 method):
\begin{itemize}
    \item For $i=1$ only the implicit step is needed. Then the explicit step performed by the \MoL results to be
        \begin{equation}
            \UU^{(1)}_*=\UU^n \qquad \UUbar^{(1)}=\zerovec
            \label{eq:MoL-step1-explicit}
        \end{equation}
        and the implicit step is
        \begin{equation}
            \UU^{(1)}=\UU^{(1)}_*+\gamma\Delta t\RR_1
            \label{eq:MoL-step1-implicit}
        \end{equation}
    \item For $i=2$ the explicit step is
        \begin{equation}
            \UU^{(2)}_*=3\UU^n-2\UU^{(1)}+\Delta t\LL_1 \qquad \UUbar^{(2)}=\dfrac{\Delta t}{2}\LL_1
            \label{eq:MoL-step2-explicit}
        \end{equation}
        and the implicit step is
        \begin{equation}
            \UU^{(2)}=\UU^{(2)}_*+\gamma\Delta t\RR_2
            \label{eq:MoL-step2-implicit}
        \end{equation}
    \item The third and final step is
        \begin{equation}
            \UU^{(3)}_*=\UU^n+\UUbar^{(2)}+\dfrac{\Delta t}{2}\LL_2
            \label{eq:MoL-step3-explicit}
        \end{equation}
        \begin{equation}
            \UU^{n+1}=\UU^{(3)}_*
            \label{eq:MoL-step3-implicit}
        \end{equation}
\end{itemize}
Notice that since $\RR_i$ is a function of $\UU^{(i)}$, the implicit step must be performed during the inversion from conservative to primitive variables by \MIR. With this scheme, we can register only the total RHS $\LL$, allowing us to utilize the existing structure of the \MoL thorn without the need for significant modifications.

\subsection{From Conserved to Primitive variables (C2P)}\label{sec:C2P}
After each \MoL step, it is necessary to recover the primitive variables from the conserved ones. Various schemes have been proposed over the years to accomplish this. Most of these schemes employ the Newton-Raphson method, which requires the partial derivatives $\pdv*{p}{\rho}$ and $\pdv*{p}{\epsilon}$. As a result, this approach performs well with analytic equations of state but may yield inaccurate results when applied to tabulated equations of state. For unmagnetized and ideal cases, however, bracketing methods have also been developed (e.g., \cite{Galeazzi2013,Kastaun2021}), which have the advantage of not requiring the computation of partial derivatives. Nevertheless, the implicit step required in the resistive case introduces significant complexity.As a result, a common choice is the use of Newton-Raphson (NR) schemes, combined with analytic equations of state (e.g., \cite{Tomei2019,Ripperda2019,Cheong2022,Mattia2023}). However, NR schemes may fail to converge or might converge to a different root if the initial guess is too far from the root or if the function behaves problematically (e.g., a zero or changing derivative). These issues can be more frequent in multidimensional schemes, where the invertibility of the Jacobian matrix is not necessarily guaranteed.

In this section, we describe the scheme implemented in the \MIR{} code, which extends the 1D scheme based on a bracketing method presented in \cite{Kastaun2021} to the resistive case.

\subsubsection{The scheme}
In the interval $\left(0,h_0^{-1}\right]$, where $h_0$ is the the lower bound for the specific enthalpy, we solve the function
\begin{equation}
    f\qty(\mu)=\mu-\dfrac{1}{\dfrac{h\qty(\mu)}{W\qty(\mu)}+W\qty(\mu)\bar{r}^2\qty(\mu)}
    \label{eq:f-c2p}
\end{equation}
where
\begin{equation}
    \bar{r}\qty(\mu)=\dfrac{M\qty(\mu)}{D}
\end{equation}
and
\begin{equation}
    \mu=\dfrac{1}{Wh}
\end{equation}
To do this, given the value of $\mu$, we perform the following steps:
\begin{enumerate}[label=\arabic*)]
    \item In the magnetized case, derive the electric field $\EE\qty(\mu)$ as described below.
    \item In the magnetized case, remove the electromagnetic counterpart from the conserved variables:
        \begin{equation}
            \Mcon\qty(\mu)=\Scon-\EE\qty(\mu)\cp\BB
        \end{equation}
        \begin{equation}
            \en_F\qty(\mu)=\en-\Uem\qty(\mu)
        \end{equation}
        \begin{equation}
            \Uem\qty(\mu)=\dfrac{B^2+E^2\qty(\mu)}{2}
        \end{equation}
        This step is performed only in the magnetized case because in the unmagnetized case one has $\Mcon=\Scon$ and $\en_F=\en$.
    \item With the hydro variables $\Mcon\qty(\mu)$ and $\en_F\qty(\mu)$ derived in the previous step, derive the hydro primitive variables
        \begin{equation}
            \rho\qty(\mu)= \max\qty{\rho_{\min},\min\qty{\hat{\rho}\qty(\mu),\rho_{\max}}}
        \end{equation}
        \begin{equation}
            \hat{\rho}\qty(\mu)=\dfrac{D}{W\qty(\mu)}
        \end{equation}        
        \begin{equation}
            \epsilon\qty(\mu)= \max\qty{\epsilon_{\min},\min\qty{\hat{\epsilon}\qty(\mu),\epsilon_{\max}}}
        \end{equation}
        \begin{equation}
            \hat{\epsilon}\qty(\mu)=W\qty(\mu)\qbar\qty(\mu)-\bar{r}\qty(\mu)\ut\qty(\mu)-1
        \end{equation}
        \begin{equation}
            \ut\qty(\mu)=\sqrt{W^2\qty(\mu)-1}
        \end{equation}
        \begin{equation}
            \qbar\qty(\mu)=\dfrac{\en_F\qty(\mu)}{D}
        \end{equation}
        \begin{equation}
            \en_F\qty(\mu)=\en-\dfrac{B^2+E^2\qty(\mu)}{2}
        \end{equation}
        \begin{equation}
            p\qty(\mu)=p\qty(\rho\qty(\mu),\epsilon\qty(\mu))
        \end{equation}
        \begin{equation}
            W\qty(\mu)=\dfrac{1}{\sqrt{1-v^2\qty(\mu)}}
        \end{equation}
        \begin{equation}
            v\qty(\mu)=\min\qty{\mu\bar{r}\qty(\mu),v_{\max}}
        \end{equation}
        Here $\rho_{\min}$ ($\rho_{\max}$) and $\epsilon_{\min}$ ($\epsilon_{\max}$) are the lower (upper) bounds for the density and the internal energy (which depend on the chosen Equation of State), respectively, and
        \begin{equation}
            v_{\max}=\min\qty{v_{\max,1},v_{\max,2}}
        \end{equation}
        with
        \begin{equation}
            v_{\max,1}=\dfrac{\ut_{\max}}{\sqrt{1+\ut_{\max}^2}}
        \end{equation}
        \begin{equation}
            v_{\max,2}=\sqrt{1-\dfrac{1}{W_{\max}^2}}
        \end{equation}
        \begin{equation}
            \ut_{\max}=\dfrac{\bar{r}_{\max}}{h_0} \qquad \bar{r}_{\max}=\dfrac{M_{\max}}{D}
        \end{equation}
        where $W_{\max}$ is the maximum allowed value for the Lorentz factor, and (see Appendix \ref{app_maxMS2Estar})
        \begin{equation}
            M_{\max}=
            \begin{cases}
                M_{\max}^{(res)} & \text{resistive case} \\
                S & \text{otherwise}
            \end{cases}
            \label{eq:M_max}
        \end{equation}
        where
        \begin{equation}
            M_{\max}^{(res)}=A_{0,\max}E_*B+\sqrt{A_{0,\max}^2E_*^2B^2+S^2}
        \end{equation}
        with $A_{0,\max}$ given by Eq. \eqref{eq:A0-max} below. The pressure is derived from $\rho$ and $\epsilon$ via an Equation of State (EoS). The list of equations implemented in \MIR{} can be found in Appendix \ref{app_eos}.
    \item With the hydro primitive variables derived in the previous step, evaluate the specific enthalpy
        \begin{equation}
            h\qty(\mu)=1+\epsilon\qty(\mu)+\dfrac{p\qty(\mu)}{\rho\qty(\mu)}
        \end{equation}
\end{enumerate}
Once we have determinate the zero of Eq. \eqref{eq:f-c2p} and the corresponding primitive variables $\qty[\rho,\epsilon,p]$, we can obtain the velocity by:
\begin{equation}
    \vv=\dfrac{\Mcon}{\rho hW^2}
    \label{eq:c2p-velocity}
\end{equation}
Finally, to ensure coherence between the electric field and the fluid velocity just obtained, the electric field is derived again, this time at a given velocity.

\subsubsection{Electric field}
In the IMHD case the electric field is given by Eq. \eqref{eq:E-imhd}, while in the resistive case it undergoes the implicit step
\begin{equation}
    \EE=\EE_*-\dfrac{W\qty[\EE+\vv\cp\BB-\qty(\vv\cdot\EE)\vv]}{\etat}
\end{equation}
with
\begin{equation}
    \etat=\dfrac{\eta/\alpha}{a_{ii}\Delta t}
    \label{eq:etat-definition}
\end{equation}
where $a_{ii}$ is the coefficient of the Runge-Kutta scheme, $\Delta t$ the time step and $\alpha$ the lapse function. The implicit step can be rewritten as
\begin{equation}
    \EE=A_0\EE_*-A_1\vv\cp\BB+A_1\qty(\vv\cdot\EE)\vv
    \label{eq:E-implicit_step}
\end{equation}
or
\begin{equation}
    \EE=A_0\EE_*-A_1\vv\cp\BB+A_2\qty(\vv\cdot\EE_*)\vv
    \label{eq:E-rmhd}
\end{equation}
where $\EE_*$ is the electric field after the \MoL step and
\begin{equation}
    A_0=\dfrac{\etat}{\etat+W}=1-A_1
    \label{eq:A0-definition}
\end{equation}
\begin{equation}
    A_1=\dfrac{W}{\etat+W}=\dfrac{1}{1+y}
    \label{eq:A1-definition}
\end{equation}
\begin{equation}
    A_2=\dfrac{W^2\etat}{\qty(1+W\etat)\qty(\etat+W)}=A_1\qty(1-\dfrac{1}{1+z})
    \label{eq:A2-definition}
\end{equation}
\begin{equation}
    y=\etat/W \qquad z=W\etat
\end{equation}
From Eq. \eqref{eq:A0-definition} it is easy to obtain the value of $A_{0,\max}$ that appears in Eq. \eqref{eq:M_max}:
\begin{equation}
    A_{0,\max}=\dfrac{\etat}{\etat+1}=1-\dfrac{1}{1+\etat}
    \label{eq:A0-max}
\end{equation}

Since the fluid velocity is given by Eq. \eqref{eq:c2p-velocity}, in step 4 we use Eq. \eqref{eq:E-imhd} (in the IMHD case) or Eq. \eqref{eq:E-rmhd} (in the resistsive case) to derive the electric field. In contrast, in step 1 the fluid velocity is not known. However, it can be written as a function of the electric field by inverting Eq. \eqref{eq:S-definition}:
\begin{equation}
    \vv=\mu\qty(\rvec-\evec\cp\bvec)
    \label{eq:velocity-inversion}
\end{equation}
where
\begin{equation}
    \rvec=\dfrac{\Scon}{D} \qquad \bvec=\dfrac{\BB}{\sqrt{D}} \qquad \evec=\dfrac{\EE}{\sqrt{D}}
\end{equation}
From Eq. \eqref{eq:E-imhd} we therefore have, in the IMHD case,
\begin{equation}
    \EE=-\mu x\rvec\cp\BB
\end{equation}
with
\begin{equation}
    x=\dfrac{1}{1+\mu b^2}
\end{equation}
In the resistive case from Eq. \eqref{eq:E-implicit_step} we can define the function
\begin{equation}
    f_i\qty(e^j)=A_0e_{*,i}-A_1\epsilon_{ilm}v^lb^m+\mu A_1\qty(r_ke^k)v_i-e_i
    \label{eq:f-electric_field}
\end{equation}
where the fluid velocity is given by Eq. \eqref{eq:velocity-inversion} and
\begin{equation}
    \evec_*=\dfrac{\EE_*}{\sqrt{D}}
\end{equation}
To derive the electric field we have to solve Eq. \eqref{eq:f-electric_field}. To do this, we use a 3D Newton-Raphson scheme. Note that since the value of $\mu$ is given, the pressure derivatives do not appear. The Jacobian matrix is
\begin{align}
    J_{ij}&=\pdv{f_i}{e^j}=\pdv{A_1}{e^j}\qty[\mu\qty(r_ke^k)v_i-e_{*,i}-\epsilon_{ilm}v^lb^m]
        \nonumber \\
    &+A_1\mu\Big[v_ir_j+\qty(r_ke^k)\pdv{v_i}{e^j}+b_ib_j -\gamma_{ij}b^2\Big]
        \nonumber \\
    & -\gamma_{ij}
\end{align}
where
\begin{equation}
    \pdv{A_1}{e^j}=\pdv{A_1}{W}\pdv{W}{e^j}=\dfrac{A_0}{\etat+W}\pdv{W}{e^j}
\end{equation}
\begin{equation}
    \pdv{W}{e^j}=\mu W^3\epsilon_{jlm}v^lb^m
\end{equation}
and
\begin{equation}
    \pdv{v_i}{e^j}=-\mu\epsilon_{ijm}b^m
\end{equation}

What has been described so far does not apply to the final \MoL step \eqref{eq:MoL-step3-implicit}. In this case, we simply have $\EE=\EE_*$.

\subsubsection{Error handling}
We now present the error policy that outlines the conditions under which unphysical values of the evolved variables are corrected, along with the methods employed for these corrections. Most of these errors arise at the stellar surfaces, where densities are low; consequently, the corrections have a minor (albeit not negligible) impact on the overall dynamics. In detail, we distinguish the following cases:

\begin{itemize}
    \item If $D<\rho_{cut}\qty(1+\delta_{\rho})$ - with $\delta_{\rho}$ a tolerance - set an artificial atmosphere (see below).
    \item If $\tau=\en-D<\tau_{atm}$, with $\tau_{atm}=\rho_{atm}\epsilon_{atm}$, adjust $\en=\tau_{atm}+D$.
    \item If $S^2>S^2_{\max}$ adjust
        \begin{equation}
            \Scon\to\Scon\sqrt{\dfrac{S^2_{\max}}{S^2}}
            \label{eq:S-adjust}
        \end{equation}
        where (see Appendix \ref{app_maxMS2Estar})
        \begin{equation}
            S^2_{\max}=\qty(\en+B\sqrt{2\en})^2
            \label{eq:S2_max}
        \end{equation}
\end{itemize}
Moreover, the C2P algorithm can occasionally return unphysical values for primitive variables. After getting the primitive variables, we check whether any correction is needed:
\begin{itemize}
    \item If $\rho<\rho_{cut}\qty(1+\delta_{\rho})$ set an artificial atmosphere (see below).
    \item In the IMHD regime, if $E\geq B$ adjust
        \begin{equation}
            \EE\to\EE\sqrt{\qty(1-\dfrac{1}{W_{\max}^2})\dfrac{B^2}{E^2}}
            \label{eq:E-adjust}
        \end{equation}
        following the idea presented in \cite{Paschalidis2013}. This adjustment is carried out only in the ideal regime because in the resistive regime it is possible to have $E>B$ \parcite{Dionysopoulou2013}.
\end{itemize}

\subsubsection{Atmosphere}\label{sec:atmosphere}
As mentioned above, we set an artificial atmosphere when $D<\rho_{cut}\qty(1+\delta_{\rho})$ or $\rho<\rho_{cut}\qty(1+\delta_{\rho})$. We enforce the density to be equal to
\begin{equation}
    \rho=\rho_{atm}\leq\rho_{cut}
    \label{eq:rho_atm}
\end{equation}
and the fluid velocity is set to zero, i.e. $\vv=\zerovec$ and $W=1$. Internal energy and pressure are evaluated via a polytropic equation of state, that is
\begin{equation}
    \epsilon=\epsilon_{atm}=K\dfrac{\rho_{atm}^{\Gamma-1}}{\Gamma-1}
    \label{eq:eps_atm}
\end{equation}
\begin{equation}
    p=K\rho_{atm}^{\Gamma}
    \label{eq:eps_atm}
\end{equation}
 where $K>0$ and $\Gamma>1$ are constants.

We implemented two different ways to evaluate the electric field. The first way consists in the IMHD equation. Since in the atmosphere we set $\vv=\zerovec$, Eq. \eqref{eq:E-imhd} implies $\EE=\zerovec$. The second way is to derive the electric field from the conserved momentum $\Scon$ following the idea presented in \cite{Paschalidis2013}. In this case the electric field results to be
\begin{equation}
    \EE=\dfrac{\BB\cp\Scon}{B^2}+\dfrac{\qty(\EE\cdot\BB)}{B^2}\BB
    \label{eq:E_from_S-general}
\end{equation}
As in \cite{Paschalidis2013} we force the electric field to be perpendicular to the magnetic field ($\EE\cdot\BB=0$), as in the IMHD and Force-Free Electrodynamics (FFE) regimes, then
\begin{equation}
    \EE=\dfrac{\BB\cp\Scon}{B^2}
    \label{eq:E_from_S}
\end{equation}
Before evaluating the electric field we must ensure that $\Scon\cdot\BB=0$. To do this, we adjust the conserved momentum as
\begin{equation}
    \Scon \to \Scon-\dfrac{\BB\cdot\Scon}{B^2}\BB
    \label{eq:S_adjusted_perp_B}
\end{equation}
Then, we must ensure that the norm of the conserved momentum is not unphysical:
\begin{equation}
    \Scon=\Scon\min\qty{1,f}
    \label{eq:S_adjusted_factor_f}
\end{equation}
where
\begin{equation}
    f=\sqrt{\qty(1-\dfrac{1}{W_{\max}^2})\dfrac{B^4}{S^2}}
    \label{eq:factor_f_for_S}
\end{equation}

\subsection{Kinematic approximation}
In \MIR{}, we have also implemented the so-called \textit{kinematic approximation}, where the fluid is assumed to be in stationary hydrodynamic equilibrium, and only electromagnetic quantities evolve over time. This approximation has been applied in works such as \cite{Bugli2014,Franceschetti2020,DelZanna2022} to study dynamo action in accretion disks and proto-neutron stars. Since all hydrodynamic quantities are predetermined, the C2P procedure is greatly simplified, as only the electric field needs to be computed. In the IMHD case, the electric field is obtained using Eq. \eqref{eq:E-imhd}, while in the RMHD case, it is computed via Eq. \eqref{eq:E-rmhd}. However, in the kinematic approximation, we can replace the evolution of the electric field with that of the momentum $\Scon$, as in \cite{Paschalidis2013}. In this approach, we first calculate $\PP=\EE\cp\BB=\Scon-\Mcon$ and then compute the electric field using Eq. \eqref{eq:E_from_S} with the substitution $\Scon \to \PP$.

\subsection{Riemann Solvers and reconstruction methods}
To solve Eq. \eqref{eq:conservatives_equations}, it is necessary to evaluate the fluxes at each cell interface. To accomplish this, a Riemann solver is required. In \MIR{} we implemented the Harten-Lax-van-Leer-Einfeldt (HLLE) solver \parcite{Harten1983}, where, for each component $i$ and each direction $j$,
\begin{equation}
    \FF^i_j=\dfrac{a_+^j\FF^i_{L,j}+a_-^j\FF^i_{R,j}-a_+^ja_-^j\qty(\UU^i_{R,j}-\UU^i_{L,j})}{a_+^j+a_-^j}
    \label{eq:HLLE}
\end{equation}
Here the subscript $R$ ($L$) means that the quantity is computed at the right (left) side of the cell interface and \parcite{DelZanna2007}
\begin{equation}
    a_{\pm}^j=\max\qty{0,\pm\lambda_{\pm,L}^j,\pm\lambda_{\pm,R}^j}
    \label{eq:a_plus_minus-HLLE}
\end{equation}
We also implemented the Lax–Friedrichs (LF) solver \parcite{Toro2009}
\begin{equation}
    \FF^i_j=\dfrac{\FF^i_{L,j}+\FF^i_{R,j}-c\qty(\UU^i_{R,j}-\UU^i_{L,j})}{2}
    \label{eq:LF}
\end{equation}
where $c=\max\qty{a_+^j,a_-^j}$, useful in case of strong jumps in pressure.

To compute the characteristic speeds $\lambda_{\pm}$ we follow the approach presented in \cite{Gammie2003}, where
\begin{equation}
    \lambda_{\pm}^i=\alpha\bar{\lambda}_{\pm}^i-\beta^i
    \label{eq:lambda_plus_minus}
\end{equation}
In the hydro and IMHD regimes we set \parcite{Gammie2003,DelZanna2007}
\begin{equation}
    \bar{\lambda}_{\pm}^i=\dfrac{
    \scriptstyle \qty(1-a^2)v^i\pm\sqrt{
    \scriptstyle a^2\qty(1-v^2)\qty[\qty(1-v^2a^2)\gamma^{ii}-\qty(1-a^2)\qty(v^i)^2]}}{1-v^2a^2}
    \label{eq:lambda_bar_plus_minus}
\end{equation}
with
\begin{equation}
    a^2=c_s^2+c_a^2-c_s^2c_a^2
    \label{eq:a2}
\end{equation}
where $c_s$ is the sound speed and $c_a$ the Alfvén speed:
\begin{equation}
    c_a^2=\dfrac{b^2}{\rho h+b^2}=\dfrac{B^2-E^2}{\rho h+B^2-E^2}
    \label{eq:ca2_speed}
\end{equation}
In the resistive case we assume $a^2=1$ as in \cite{DelZanna2007} and \cite{Bucciantini2013}, so
\begin{equation}
    \lambda_{\pm}^i=\pm\alpha\sqrt{\gamma^{ii}}-\beta^i
\end{equation}
In order to compute the values at each side of cell's interfaces we adopt the weno-z scheme \parcite{Borges2008}. However, we also implemented the second–order total variation diminishing (TVD) \textit{minmod} and \textit{monotonized central} (MC2) methods \parcite{Toro2009}.

\subsection{Divergence of electric field}
To evaluate the electric current given by Eq. \eqref{eq:J-vector}, we need to determine the electric charge, which is provided by Eq. \eqref{eq:Maxwell1}, specifically the divergence of the electric field. We have implemented two different methods to compute this. The first method employs finite differences. The second method utilizes the Riemann solver to compute the divergence. In this case, the flux is determined by the electric field and can be calculated at each cell interface using Eq. \eqref{eq:HLLE} with $\UU^i_{R,j}=\UU^i_{L,j}=0$ (see Appendix \ref{app_divergence} for the proof).

%%%%%%%%%%%%%%%%%%%%%%%%%%%%%%%%%%%%%%%%%%%%%%%%%%%%%%%
\section{Numerical results} \label{sec4}
As is well known, the validation of a numerical code is a crucial step to ensure the reliability of the results obtained and its applicability to real-world problems.

In this section, we present the results obtained from the tests conducted on our code. These tests, already documented in the literature (e.g., \cite{Mosta2014,DelZanna2007,Bucciantini2013,Cipolletta2020}), were designed to evaluate the accuracy, efficiency, and robustness of the code in both the hydro regime and the MHD regime, both ideal and resistive. Additionally, these tests aim to evaluate our C2P scheme and investigate the impact of artificial dissipation (necessary in the ideal regime to prevent spurious oscillations), specifically examining its necessity within the resistive regime.

Finally, we present the results of numerical simulations performed in the presence of a bar-mode instability (e.g. \cite{Franci2013,DePietri2014}), a rotational instability that occurs in highly dense and rapidly rotating astrophysical objects, such as newly formed neutron stars or black holes. This instability is named after the bar-like or elongated shape that the rotating object can assume when it loses symmetry. The aim of these simulations, performed in both the ideal and resistive regimes, is to assess the code's capability to handle non-linear instabilities and highly dynamic scenarios.

The time evolution of the metric was carried out using the BSSN evolution system. As usual, artificial dissipation was applied, with a dissipation parameter set to 0.1. Simulations in non-resistive regimes were performed using either the second-order (RK2) or fourth-order (RK4) Runge-Kutta time integrators, with the CFL factor set to 0.125 or 0.25, respectively. In contrast, simulations in the resistive regime were performed using the SSP2(2,2,2) time inetragtor with a CFL factor of 0.125.

\begin{figure}
    \centering
    \includegraphics[width=0.45\textwidth]{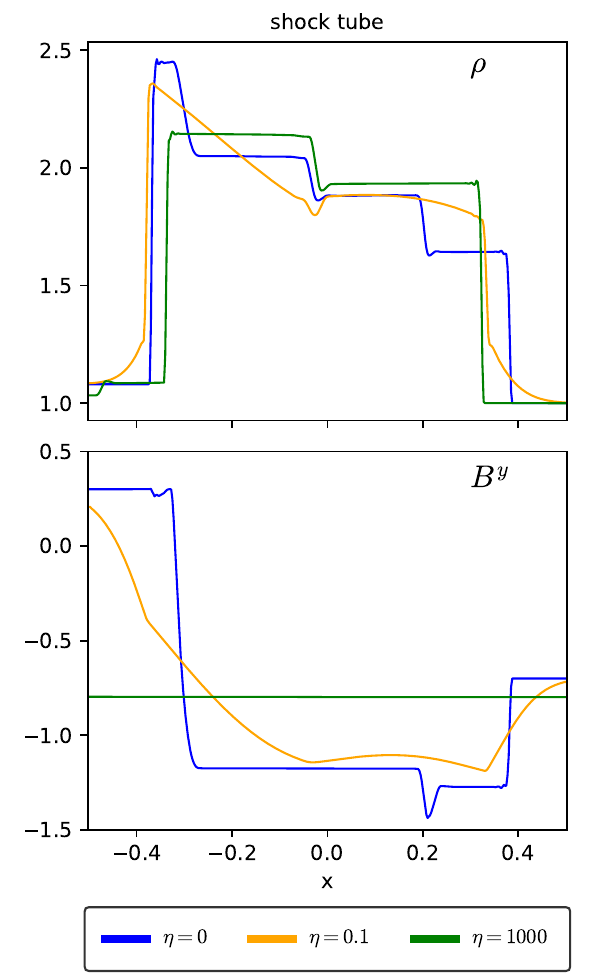}
    \caption{Results of the shock tube problem at the final time $t=0.55$. The upper panel shows the density, while the lower panel the $y-$component of the magnetic field. The blue line is the case $\eta=0$, the orange line is the case $\eta=0.1$, while the green line is the case $\eta=1000$.}
    \label{fig:shock_tube}
\end{figure}%

\subsection{Shock Tube}
The first test is the shock tube presented in \parcite{Dumbser2009}, performed in both IMHD and resistive regimes. As in \cite{Bucciantini2013}, the test is performed in a stationary Cartesian grid (i.e. Minkowski spacetime) using the Lax-Friedrichs flux solver, and the TVD MC2 reconstruction method. The RK2 scheme is employed as the time integrator in the IMHD case. The electrical resistivity is constant throughout the domain. The initial conditions are
\begin{multline}
    \qty(\rho,p,v^x,v^y,v^z,B^x,B^y,B^z)=\\
    \qty(1.08, 0.95, 0.4, 0.3, 0.2, 2.0, 0.3, 0.3) \text{ for } x<0
\end{multline}
and
\begin{multline}
    \qty(\rho,p,v^x,v^y,v^z,B^x,B^y,B^z)=\\
    \qty(1.0,1.0,-0.45,-0.2,0.2,2.0,-0.7,0.5) \text{ for } x>0
\end{multline}
while the initial electric field is set equal to the IMHD value. The $x-$component spans over the range $\qty[-15,15]$. The grid is uniform with 400 cells. The final time is $t=0.55$. The test was repeated for three different values of electrical resistivity: $\eta=0$, $\eta=0.1$, and $\eta=1000$. The adiabatic index of the Ideal Fluid EoS is $\Gamma=5/3$. Results are shown in Figure \ref{fig:shock_tube}. We observe an exact match with those presented in \cite{Bucciantini2013} for all cases.

\subsection{Self-similar current sheet}
This problem was first proposed by \cite{Komissarov2007}. Following the same approach, it was later presented by \cite{Palenzuela2009}, \cite{Dumbser2009}, and \cite{Bucciantini2013}. The exact solution for the $y$-component of the magnetic field in the limit of infinite pressure is 
\begin{equation}
    B^y(x,t;\eta)=B_0\erf\qty(\dfrac{x}{2\sqrt{\eta t}})
    \label{eq:current_sheet_analytical}
\end{equation}
where $\erf$ is the error function. The initial conditions are $\rho=1$, $p=50$, $\vv=\zerovec$, $\EE=\zerovec$, $\BB=\qty(0,B^y\qty(x,t_i;\eta),0)$, being $t_i$ the initial time. As in \cite{Bucciantini2013}, we have adopted $B_0=1$ and an adiabatic coefficient $\Gamma=4/3$. The computational domain spans the range $x\in\qty[-1.5,1.5]$ and is structured using a uniform grid consisting of 200 cells. In \cite{Komissarov2007} and subsequent works, simulation results were presented for $\eta=10^{-2}$, $t_i=1$, and $t_f=10$ (where $t_f$ denotes the final time). In this work, we report the results obtained with four different values of $\eta$, while keeping the initial and final products $\eta t_i$ and $\eta t_f$ constant. Specifically, the values of $t_i$ and $t_f$ were chosen based on the value of $\eta$ such that $\eta t_i=0.01$ and $\eta t_f=0.1$. The four triplets used are therefore
\begin{equation}
    \qty(\eta,t_i,t_f)=\qty(10^{-1},0.1,1)
    \label{eq:current_sheet_eta-1}
\end{equation}
\begin{equation}
    \qty(\eta,t_i,t_f)=\qty(10^{-2},1,10)
    \label{eq:current_sheet_eta-2}
\end{equation}
\begin{equation}
    \qty(\eta,t_i,t_f)=\qty(10^{-3},10,100)
    \label{eq:current_sheet_eta-3}
\end{equation}
\begin{equation}
    \qty(\eta,t_i,t_f)=\qty(10^{-4},100,1000)
    \label{eq:current_sheet_eta-4}
\end{equation}
Despite the evolution is predominantly resistive (in the limit of infinite pressure, only the magnetic field evolves), the problem is addressed in the fully dynamical regime using the RK2 scheme in the ideal case, as in \cite{Bucciantini2013}.

\begin{figure}
    \centering
    \includegraphics[width=0.45\textwidth]{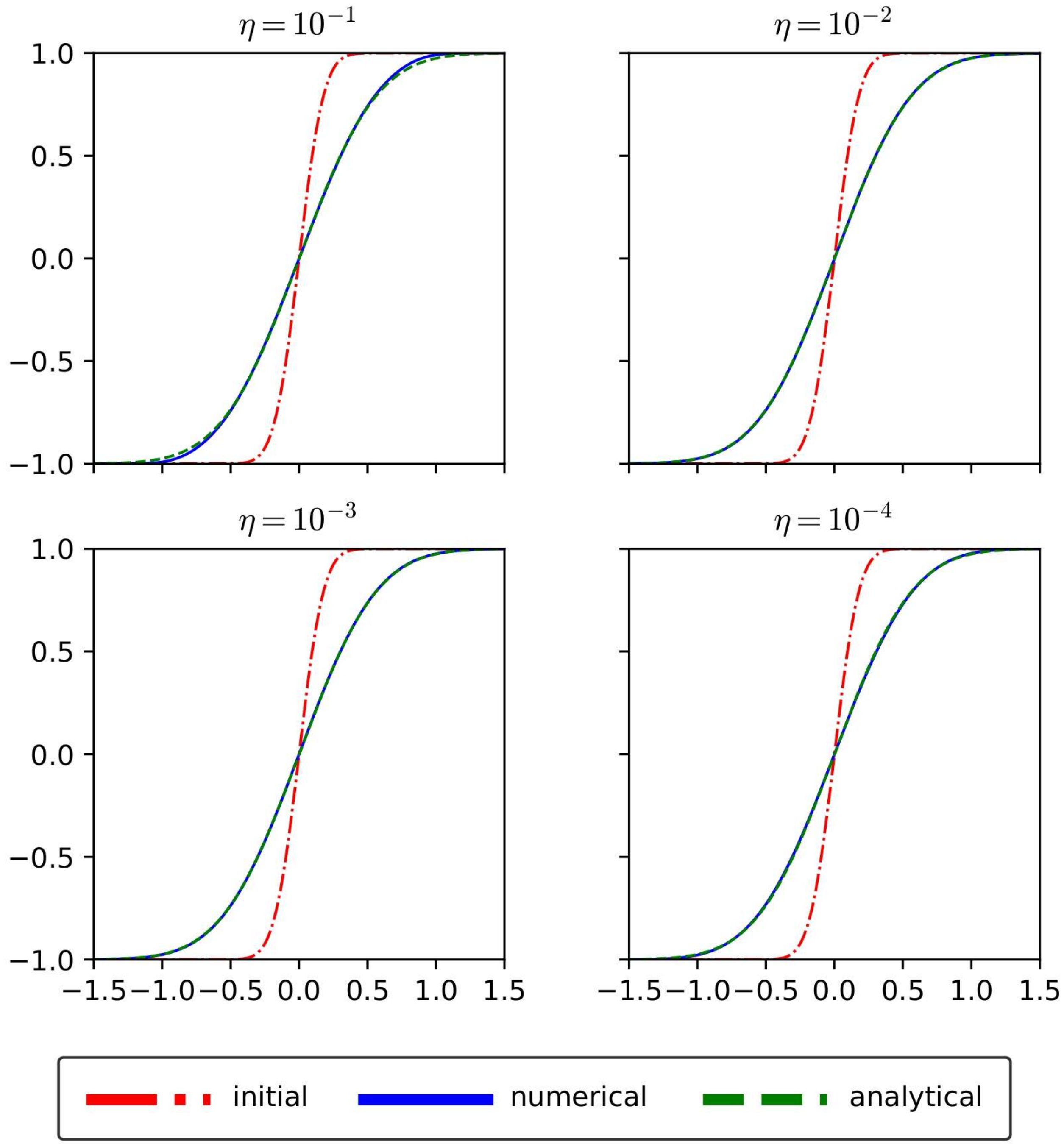}
    \caption{Evolution of the $y-$component of the magnetic field in a self-similar current sheet for different electrical resistivity values $\eta$. The red dot-dashed line represents the initial condition. The blue solid line depicts the numerical solution, which is indistinguishable from the green dashed line representing the exact solution given by Eq. \eqref{eq:current_sheet_analytical}. The initial and final times for each simulation are detailed in the text (Eqs. \eqref{eq:current_sheet_eta-1}-\eqref{eq:current_sheet_eta-4}).}
    \label{fig:current_sheet}
\end{figure}

By keeping the spatial numerical grid unchanged, we are able to evaluate the impact of numerical viscosity on our simulations. The initial and final configurations are theoretically identical regardless of the value of $\eta$, while the intermediate evolution depends on the electrical resistivity. Significant discrepancies between the analytical and numerical solutions would indicate a considerable influence of numerical viscosity. However, as illustrated in Figure \ref{fig:current_sheet}, there is excellent agreement between the two solutions (the numerical solution is shown in blue, the analytical solution in green, and the red line represents the initial configuration). This demonstrates that, at this resolution, the intrinsic resistivity of the numerical scheme is less than $10^{-4}$.

\begin{figure}
    \centering
    \includegraphics[width=0.45\textwidth]{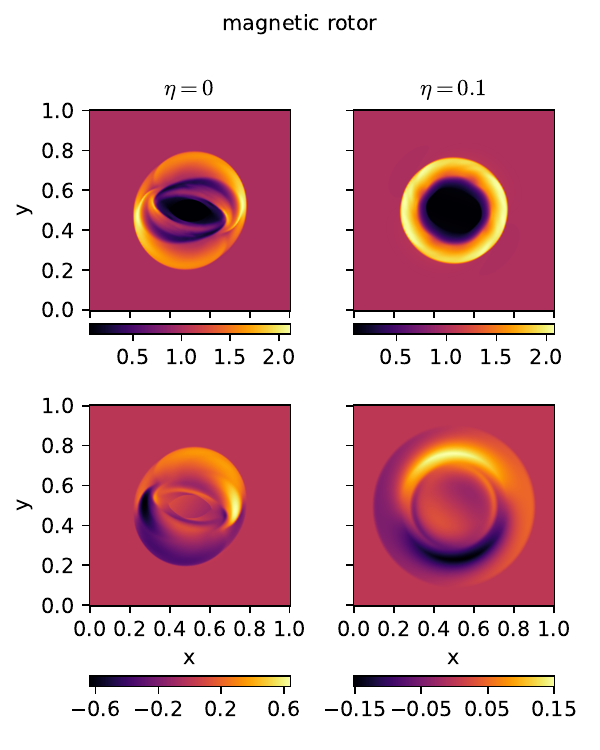}
    \caption{Results of the magnetized rotor problem at the final time $t=0.3$. The upper panels show the pressure, while the lower panels the $z-$component of the electric field. The left column is the IMHD case $\eta=0$, while the right column is the resistive case $\eta=0.1$.}
    \label{fig:test-rotor}
\end{figure}

\subsection{Magnetic rotor}
We now present the results of a standard 2D test. The relativistic IMHD version was first proposed by \cite{DelZanna2003}, while the resistive version in \cite{Dumbser2009}. The initial configuration is a circular region with radius $r=0.1$, density $\rho=10$, rotating with a uniform angular velocity $\Omega=8.5$. The region is located in a static medium with density $\rho=1$. Uniform pressure $p=1$ and magnetic field $\qty(B^x,B^y,B^z)=\qty(1,0,0)$ are present in the whole domain. The adiabatic index of the Ideal Fluid EoS is $\Gamma=4/3$. The initial electric field is set equal to the IMHD value. The final time is $t=0.3$. The computational domain is $x=\qty[0,1]$ and $y=\qty[0,1]$, and the center of the circular region is located at $\qty(x,y)=\qty(0.5,0.5)$. The grid is uniform with $400\times 400$ cells. The test was repeated in the IMHD regime and in the resistive regime with $\eta=0.1$. As in \cite{Bucciantini2013}, the test was performed in a stationary Cartesian grid, using the Lax-Friedrichs flux solver, and the TVD MC2 reconstruction method. Also in this case, the RK2 scheme is employed as the time integrator in the IMHD case. The electrical resistivity is set constant throughout the entire domain. Results are shown in Figure \ref{fig:test-rotor}. We observe an exact match with those presented in \cite{Bucciantini2013} for both cases.

\subsection{TOV star}
\begin{figure}
    \centering
    \includegraphics[width=0.45\textwidth]{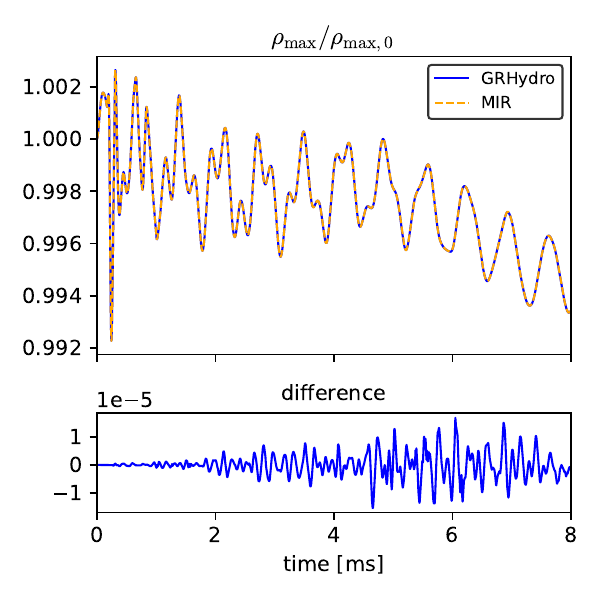}
    \vspace{-5mm}
    \caption{Result of the TOV star simulation. Top: time evolution of the maximum rest-mass density $\rho_{\max}$ normalized to the value at the initial time $\rho_{\max,0}$. Bottom: time evolution of the difference $\qty[\rho_{\max \text{(GRHydro)}}-\rho_{\max \text{(MIR)}}]/\rho_{\max,0}$.}
    \label{fig:tov_GRHydro_vs_MIR}
\end{figure}

To assess the stability and the accuracy of our code, we consider the evolution of a non-rotating stable star. The initial configuration was built with the \texttt{RNS} code \parcite{Font2000} using a polytropic star with adiabatic index $\Gamma=2$, polytropic constant $K=100$, and initial central rest-mass density $\rho_c=\SI{1.28e-3}{}$. We performed the evolution adopting the Ideal Fluid EoS with the same value of $\Gamma$. The test is run on a cubic Cartesian grid with 4 refinements levels. The $x-$, $y-$ and $z-$coordinates of the finer grid span over the range $\qty[-15,15]$ with grid steps $\dd x=\dd y=\dd z=0.375$. The test is simulated for $\SI{8}{ms}$ using the weno-z reconstruction method, the HLLE flux solver, and the RK4 time integrator. To investigate the accuracy of our code, we compare the result of the test with that obtained using the \texttt{GRHydro} code. Figure \ref{fig:tov_GRHydro_vs_MIR} illustrates the time evolution of the maximum rest-mass density $\rho_{\max}$, showing a good match.

\begin{figure}
    \centering
    \includegraphics[width=0.45\textwidth]{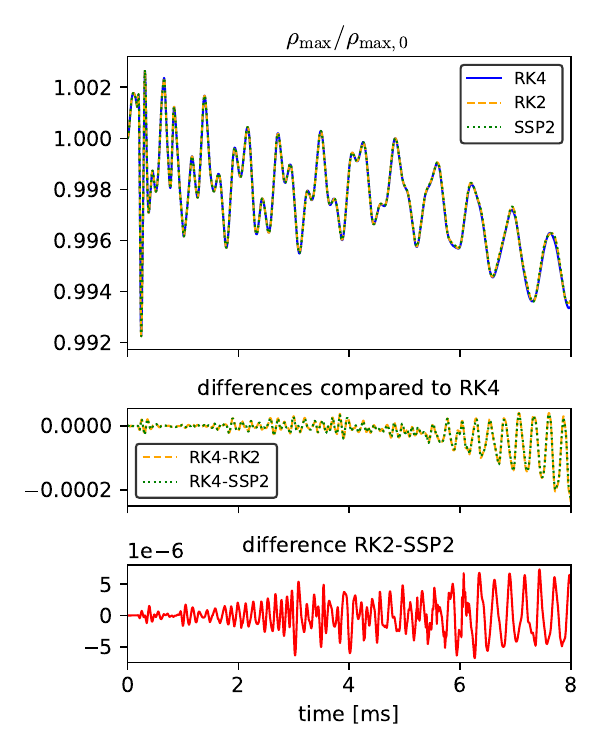}
    \vspace{-5mm}
    \caption{Result of the TOV star simulations performed with the GRHydro code. Top: time evolution of the maximum rest-mass density $\rho_{\max}$ normalized to the value at the initial time $\rho_{\max,0}$. Center: time evolution of the differences compared to the RK4 scheme. Bottom: time evolution of the difference between the RK2 and SSP2(2,2,2) schemes.}
    \label{fig:tov_rk2_vs_rk4}
\end{figure}

The TOV test was additionally employed to verify the correct implementation of the SSP2(2,2,2) scheme within the \MoL thorn. Simulation results obtained using the GRHydro code with the RK4 scheme were compared to those obtained with the RK2 and SSP2(2,2,2) schemes. A correct implementation of the SSP2(2,2,2) scheme in the MOL thorn was expected to reproduce the same results as simulations conducted with the other RK schemes. The good agreement among the results of the various simulations, as shown in Figure \ref{fig:tov_rk2_vs_rk4}, provides further confirmation - along with the agreement of the discussed tests above with those reported in the literature - of the correct implementation of the SSP2(2,2,2) scheme in the \MoL thorn.

\begin{figure}
    \centering
    \includegraphics[width=0.45\textwidth]{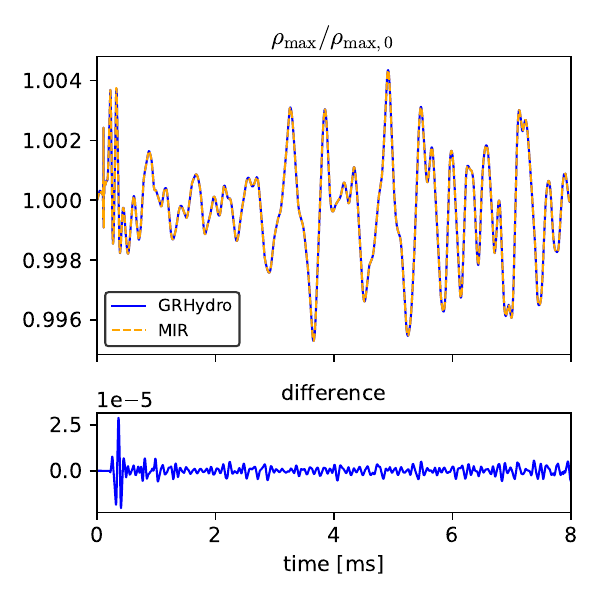}
    \vspace{-5mm}
    \caption{Result of the magnetized rotating star simulation. Top: time evolution of the maximum rest-mass density $\rho_{\max}$ normalized to the value at the initial time $\rho_{\max,0}$. Bottom: time evolution of the difference $\qty[\rho_{\max \text{(GRHydro)}}-\rho_{\max \text{(MIR)}}]/\rho_{\max,0}$.}
    \label{fig:utst_GRHydro_vs_MIR}
\end{figure}

\subsection{Magnetized rotating star}
In order to test the C2P scheme for the magnetized case, we consider the evolution of a magnetized and uniformly rotating stable star. The initial configuration was built with the \texttt{XNS} code \parcite{Bucciantini2011,Pili2014} using a polytropic star with adiabatic index $\Gamma=2$, polytropic constant $K=100$, initial central rest-mass density $\rho_c=\SI{1.28e-3}{}$, rotation rate $\Omega=\SI{1.1e-2}{}$, and poloidal magnetic field with maximum of $B_{\max}\approx\SI{2.63e12}{Gauss}$. The test is run on a cubic Cartesian grid with 4 refinement levels. The $x-$, $y-$ and $z-$coordinates of the finer grid span over the range $\qty[-15,15]$ with grid steps $\dd x=\dd y=\dd z=0.25$. The test is simulated for $\SI{8}{ms}$ using the weno-z reconstruction method, the HLLE flux solver, and the fourth-order Runge-Kutta (RK4) time integrator with a CFL factor of $0.25$. To remove spurious oscillations in the magnetic field we add a fifth-order Kreiss-Oliger artificial dissipation \parcite{Kreiss1973} to the magnetic field evolution equation with a dissipation parameter of $10^{-2}$. Furthermore, to compare the simulation result with that obtained using the \texttt{GRHydro} code, we also enforced the magnetic field in the atmosphere to be zero in \MIR{}. The initial electric field is determined by the IMHD relation, and is evolved using the same prescription. Figure \ref{fig:utst_GRHydro_vs_MIR} illustrates the time evolution of the maximum rest-mass density $\rho_{\max}$, also showing a good match in this case.

\begin{figure}
    \centering
    \includegraphics[width=0.45\textwidth]{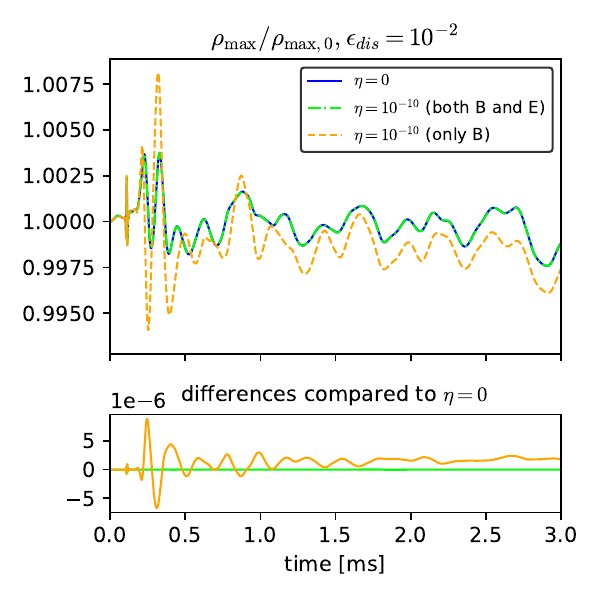}
    \vspace{-5mm}
    \caption{Result of the magnetized rotating star simulation in the IMHD and resistive regimes. The dissipation parameter is $\epsilon=10^{-2}$. Top: time evolution of the maximum rest-mass density $\rho_{\max}$ normalized to the value at the initial time $\rho_{\max,0}$. Blue solid line is the ideal regime. Green dashed line is the resistive regime with artificial dissipation added to both magnetic and electric field. Orange dotted line is the resistive regime with artificial dissipation added only to the magnetic field. Bottom: time evolution of the difference $\qty[\rho_{\max (\eta=0)}-\rho_{\max (\eta=10^{-10})}]/\rho_{\max,0}$. Colors are the same as the top panel.}
    \label{fig:utst_imhd_vs_res1m10_dis1m2_rho_3ms}
\end{figure}

\begin{figure}
    \centering
    \includegraphics[width=0.45\textwidth]{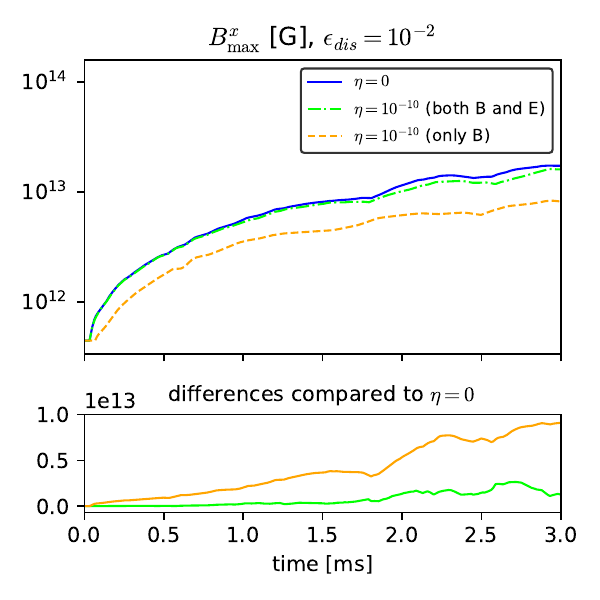}
    \vspace{-5mm}
    \caption{Result of the magnetized rotating star simulation in the IMHD and resistive regimes. The dissipation parameter is $\epsilon=10^{-2}$. Top: time evolution of the $x-$component of the magnetic field in base-10 log scale. Bottom: time evolution of the difference $B^x_{(\eta=0)}$-$B^x_{(\eta=10^{-10})}$. Colors are the same as in Figure \ref{fig:utst_imhd_vs_res1m10_dis1m2_rho_3ms}.}
    \label{fig:utst_imhd_vs_res1m10_dis1m2_Bx_3ms}
\end{figure}

\begin{figure}
    \centering
    \includegraphics[width=0.45\textwidth]{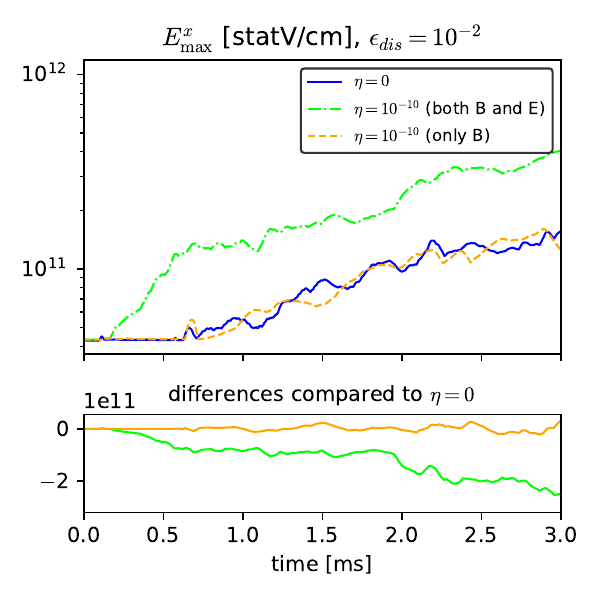}
    \vspace{-5mm}
    \caption{The same thing as Figure \ref{fig:utst_imhd_vs_res1m10_dis1m2_Bx_3ms}, but for the $x-$component of the electric field.}
    \label{fig:utst_imhd_vs_res1m10_dis1m2_Ex_3ms}
\end{figure}

\subsection{Resistive magnetized rotating star}
We now consider the evolution of a magnetized and uniformly rotating star in the resistive regime. The initial configuration is exactly the same as the previous test. However, the evolution is now performed using the SSP2(2,2,2) scheme with a CFL factor of $0.125$. The electrical resistivity is constant within the star and zero outside, where the electric field is evaluated using the IMHD relation. Figure \ref{fig:utst_imhd_vs_res1m10_dis1m2_rho_3ms} illustrates the time evolution of the maximum rest-mass density $\rho_{\max}$, normalized to its initial value $\rho_{\max,0}$, in both ideal and resistive regimes. The electrical resistivity is set to $\eta=10^{-10}$, and the dissipation parameter is $\epsilon_{dis}=10^{-2}$. Artificial dissipation is added only to the magnetic field evolution equation in the ideal regime (blue solid line) and in the resistive simulation represented by the orange dotted line. In contrast, it is also included in the electric field evolution equation for the simulation represented by the green dashed line. We observe a good match between the rest-mass density in the ideal regime and that in the resistive regime when artificial dissipation is applied to both magnetic and electric fields. In contrast, a significant difference is noted in the resistive regime when artificial dissipation is added solely to the magnetic field. Similar observations can be made for the magnetic field (Figure \ref{fig:utst_imhd_vs_res1m10_dis1m2_Bx_3ms}), while the opposite effect is observed for the electric field (Figure \ref{fig:utst_imhd_vs_res1m10_dis1m2_Ex_3ms}). This discrepancy is likely due to the dissipation term having the same sign as the flux term of the electric field and an opposite sign to that of the magnetic field.

\begin{figure}
    \centering
    \includegraphics[width=0.45\textwidth]{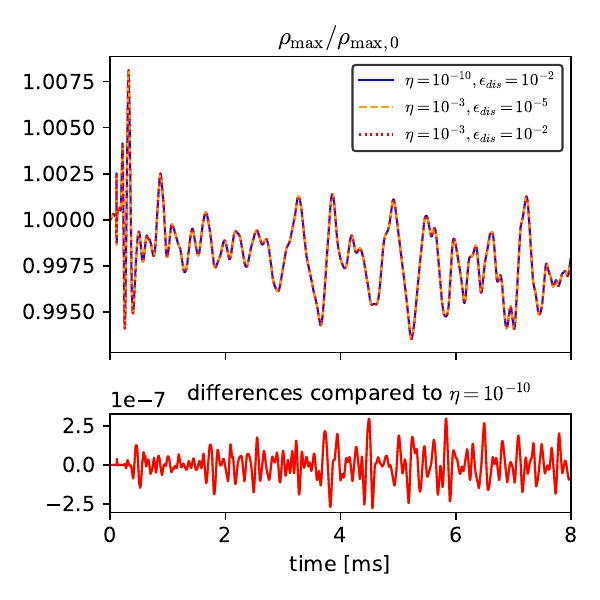}
    \vspace{-5mm}
    \caption{Result of the magnetized rotating star simulation in the resistive regime for different values of $\eta$ and $\epsilon_{dis}$, with artificial dissipation added only to the magnetic field. Top: time evolution of the maximum rest-mass density $\rho_{\max}$ normalized to the value at the initial time $\rho_{\max,0}$. Blue solid line and red dotted line correspond to simulations with the same dissipation parameter but different electrical resistivity, while orange dashed line and red dotted line correspond to simulations with the same electrical resistivity but different dissipation parameter. Bottom: time evolution of the difference $\qty[\rho_{\max (\eta=0)}-\rho_{\max (\eta=10^{-10})}]/\rho_{\max,0}$.  Colors are the same as the top panel.}
    \label{fig:utst_resistive_rho_8ms}
\end{figure}

\begin{figure}
    \centering
    \includegraphics[width=0.45\textwidth]{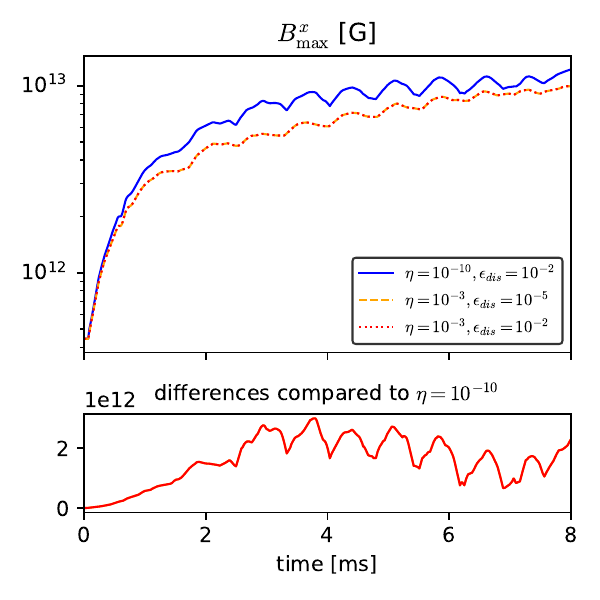}
    \vspace{-5mm}
    \caption{Result of the magnetized rotating star simulation in the resistive regime for different values of $\eta$ and $\epsilon_{dis}$, with artificial dissipation added only to the magnetic field. Top: time evolution of the $x-$component of the magnetic field in base-10 log scale. Bottom: time evolution of the difference $B^x_{(\eta=10^{-10})}$-$B^x_{(\eta=10^{-3})}$. Colors are the same as in Figure \ref{fig:utst_resistive_rho_8ms}.}
    \label{fig:utst_resistive_Bx_8ms}.
\end{figure}

\begin{figure}
    \centering
    \includegraphics[width=0.45\textwidth]{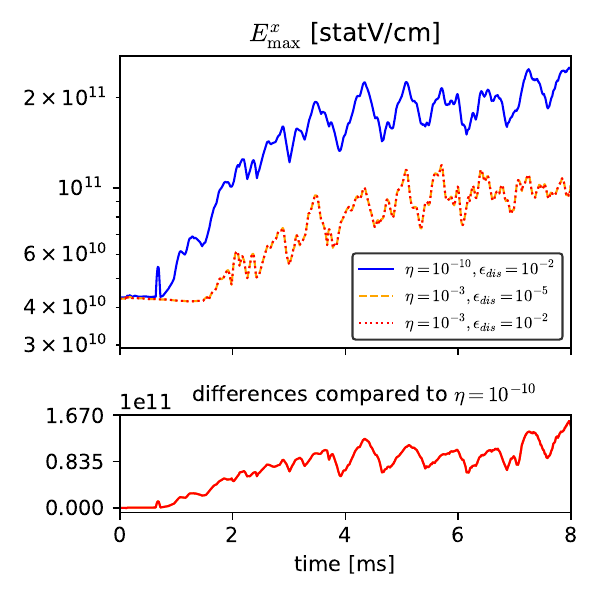}
    \vspace{-5mm}
    \caption{The same thing as Figure \ref{fig:utst_resistive_Bx_8ms}, but for the $x-$component of the electric field.}
    \label{fig:utst_resistive_Ex_8ms}
\end{figure}

To determine the dominant term between the resistive contribution and the artificial dissipation term, we repeated the simulation by varying the values of $\eta$ and $\epsilon_{dis}$, applying the artificial dissipation solely to the magnetic field. The results are illustrated in Figures \ref{fig:utst_resistive_rho_8ms}-\ref{fig:utst_resistive_Ex_8ms} for the cases $\qty(\eta,\epsilon_{{dis}})=\qty(10^{-10},10^{-2}),\qty(10^{-3},10^{-5}),\qty(10^{-3},10^{-2})$. Concerning the density, a good agreement is observed among the three cases. For the electromagnetic fields, a consistent match is noted between simulations with the same electrical resistivity, regardless of the artificial dissipation parameter value. Thus, the difference with the case $\eta=10^{-10}$ can be attributed solely to electrical resistivity.

\subsection{Bar-mod instability}
With the successful validation of our code and the demonstration that artificial dissipation is redundant in the resistive regime, we are now able to present, to our knowledge, the first simulation results of bar-mode instability in a resistive regime.

The bar-mode instability is a physical process that occurs in rapidly rotating astrophysical bodies, such as neutron stars or other compact stellar systems. This instability develops when a self-gravitating object rotates at a sufficiently high rate to experience non-spherical deformations. In particular, bar-mode instability describes a scenario in which an initially axisymmetric configuration becomes unstable, evolving into an ellipsoidal or "bar-like" structure instead of preserving its original symmetry. This deformation alters the mass quadrupole moment of the system, leading to the emission of gravitational waves. Consequently, bar-mode instability is of significant interest in the study of gravitational wave sources, as it can provide valuable insights into the internal composition and rotational dynamics of neutron stars and other compact rotating objects. Moreover, by breaking axial symmetry, this instability can create conditions conducive to the development of a dynamo or magnetorotational instability (MRI).

The initial configuration, built with the \texttt{XNS} code, is based on the magnetized U13 model presented in \cite{Franci2013}, an unstable configuration that extends the sequence of models presented in \cite{Stergioulas2004} and \cite{Dimmelmeier2006}. It consists of a differentially rotating star with central density $\rho_c=\SI{0.599e-4}{}$ and polar-to-equatorial ratio $r_p/r_e=0.200$. The initial toroidal vector potential $A_{\phi}$, used to perturb the equilibrium with a poloidal magnetic field, is given by
\begin{equation}
    A_{\phi}=A_b\qty(\max\qty{p-p_{cut},0})^2
\end{equation}
where $p_{cut}$ is $4\%$ of the maximum pressure. Here $A_b$ is chosen to have a maximum magnetic field of the order of $10^9$ Gauss. The final time is $\SI{25}{ms}$, and four different values for electrical resistivity in the resistive regime were used: $10^{-12}$, $10^{-9}$, $10^{-6}$, and $10^{-3}$. The lowest value corresponds to
\begin{equation}
    \eta=10^{-12}=\SI{5.0e-18}{s}=\SI{5.6e-7}{\Omega.m}
\end{equation}
while the value $\eta=10^{-6}$ corresponds to the electrical resistivity at the base of the lower solar atmosphere \parcite{Chae2021}. In the ideal regime, two different values for the dissipation parameter $\epsilon_{dis}$ were used: $0.1$ (as in \cite{Franci2013}) and $0.05$, while in the resistive regime no artificial dissipation was added.

\begin{figure}
    \centering
    \includegraphics[width=0.45\textwidth]{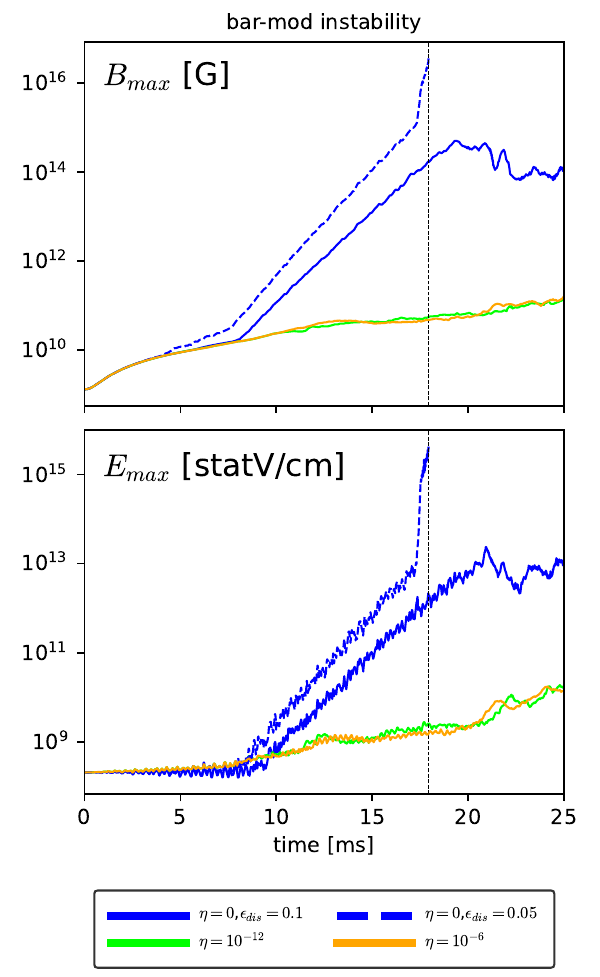}
    \vspace{-5mm}
    \caption{Time evolution of the maximum of the magnetic field (top panel) and electric field (bottom panel) in base-10 log scale for the bar-instability simulation in the IMHD and resistive regimes. The black dotted vertical line indicates the time when the simulation $\qty(\eta,\epsilon_{dis})=\qty(0,0.05)$ failed.}
    \label{fig:u13_imhd_vs_res_norms}.
\end{figure}
\begin{figure*}
    \centering
    \includegraphics[width=0.90\textwidth]{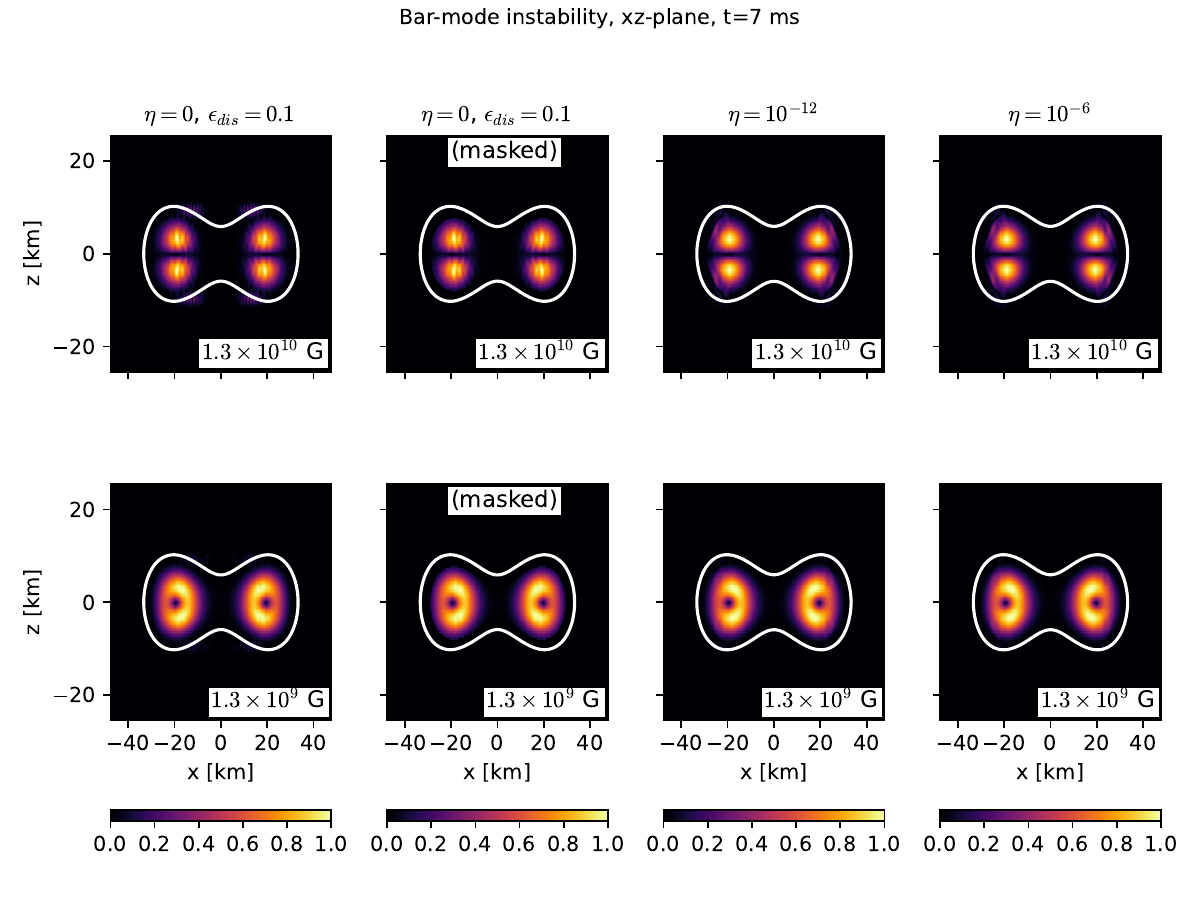}
    \vspace{-5mm}
    \caption{Spatial distribution, on the $xz$-plane at $t=\SI{7}{ms}$, of the norm of the magnetic field (top panel) and its poloidal component (bottom panel), for the IMHD simulation with $\epsilon_{dis}=0.1$ (first and second column), and two resistive simulations: $\eta=10^{-12}$ (third column) and $\eta=10^{-6}$ (fourth column). The second column corresponds to the ideal simulation in which the surface has been masked, in order to show the distribution inside the star.}
    \label{fig:u13_imhd_vs_res_xz_7ms}
\end{figure*}
\begin{figure*}
    \centering
    \includegraphics[width=0.90\textwidth]{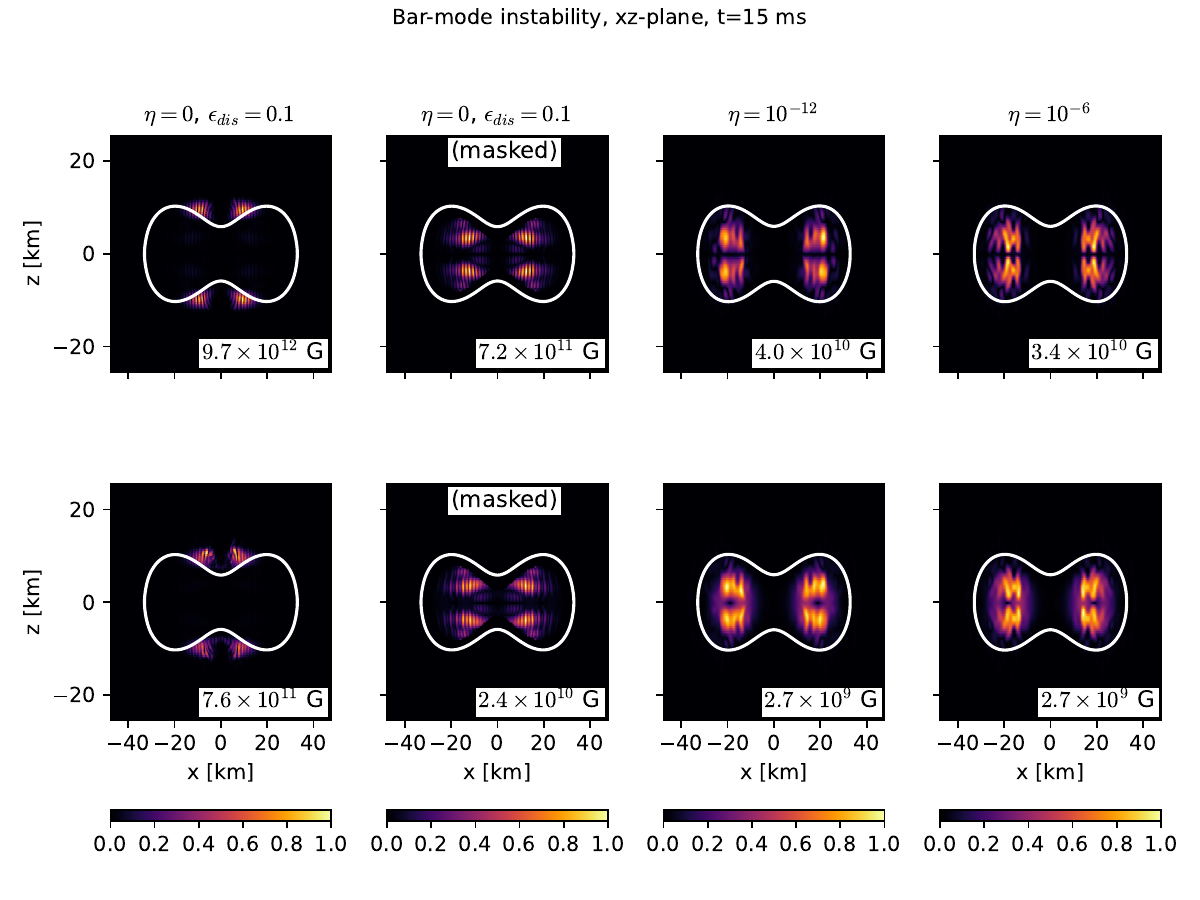}
    \vspace{-8mm}
    \caption{The same plot as in Figure \ref{fig:u13_imhd_vs_res_xz_7ms}, but at $t=\SI{15}{ms}$. After the development of the bar-mode instability, one sees that, quite expected, there is a clear distinction between the Ideal and the resistive simulations. We would argue in the main text how this is related to the effect of using artificial dissipation that seems unable to suppress physical effects to numerical ones.}
    \label{fig:u13_imhd_vs_res_xz_15ms}
\end{figure*}

As it is possible to seen in Figure \ref{fig:u13_imhd_vs_res_norms}, where the time evolution of the maximum of the magnetic and electric field is shown in log-10 base scale, the simulation with $\epsilon_{dis}=0.05$ (blue dashed line) failed after approximately $\SI{18}{ms}$ (black dotted vertical line). This was due to the generation of the magnetic field on the star's surface caused not by physical phenomena but by numerical errors. The same thing happens for $\epsilon_{dis}=0.1$ (blue solid line), but in this case the artificial dissipation is sufficient to prevent the failure. This phenomenon can be observed in the first columns of Figures \ref{fig:u13_imhd_vs_res_xz_7ms} and \ref{fig:u13_imhd_vs_res_xz_15ms}, where the spatial distribution on the $xz$-plane of the magnetic field (top panel) and its poloidal component (bottom panel; defined as in \cite{Franci2013}) is shown at two different times. At $\SI{7}{ms}$ (Figure \ref{fig:u13_imhd_vs_res_xz_7ms}), the maximum magnetic field is still situated within the star; however, magnetic field generation begins to occur at the stellar surface. Subsequently, an exponential growth of the magnetic field begins, with its peak eventually aligning with the structures located near the surface (Figure \ref{fig:u13_imhd_vs_res_xz_15ms}). The emergence and development of these structures are likely attributed to the imposition of a zero magnetic field in the artificial atmosphere. Consequently, the discontinuity of the magnetic field at the surface introduces errors that act as seeds for non-physical instabilities.

\begin{figure}
    \centering
    \includegraphics[width=0.45\textwidth]{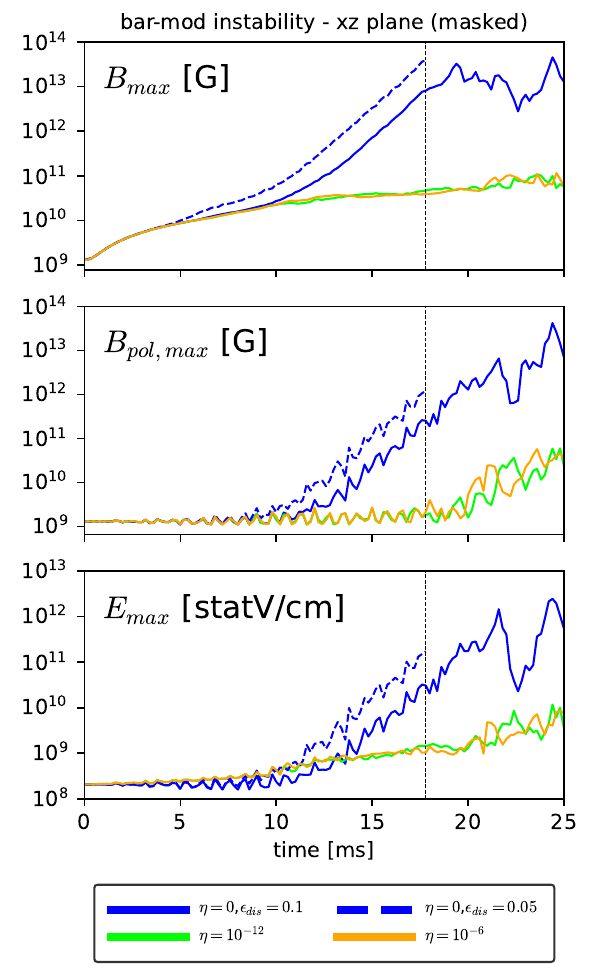}
    \vspace{-2mm}
    \caption{Time evolution of the maximum of the magnetic field (top panel), its poloidal component (central panel), and the electric field (bottom panel) in base-10 log scale for the bar-instability simulation in the IMHD and resistive regimes, on the $xz$-plane. The data correspond to the values in the inner region of the star, i.e. the surface region has been masked. The black dotted vertical line indicates the time when the simulation $\qty(\eta,\epsilon_{dis})=\qty(0,0.05)$ failed.}
    \label{fig:u13_imhd_vs_res_norms_xz}.
\end{figure}

The same exponential growth also occurs in the internal regions of the star, as shown in Figure \ref{fig:u13_imhd_vs_res_norms_xz}. Here, the growth is caused by the bar-mode instability; however, it is still dependent on the value of the artificial dissipation. Furthermore, the electric field remains nearly constant until about $\SI{8}{ms}$, which coincides with the beginning of exponential growth. This is because the poloidal component also remains constant until that point. This indicates that the initial growth of the magnetic field is exclusively related to the creation and growth of the toroidal component.

A non-zero electrical resistivity significantly alters the dynamics of the system. As shown in Figure \ref{fig:u13_imhd_vs_res_norms}, which reports the results of the simulations with $\eta=10^{-12}$ (lime solid line) and $\eta=10^{-6}$ (orange solid line), the magnetic and electric fields grow much more slowly compared to the ideal regime, without exhibiting the drastic exponential growth after about $\SI{8}{ms}$. The absence of numerical instabilities near the stellar surface (third and fourth columns in Figures \ref{fig:u13_imhd_vs_res_xz_7ms} and \ref{fig:u13_imhd_vs_res_xz_15ms}) implies that the field maxima (whose time evolution is shown in Figures \ref{fig:u13_imhd_vs_res_norms} and \ref{fig:u13_imhd_vs_res_norms_xz}) are located within the star. Moreover, all the figures show virtually no difference between the two resistive regimes. The same results were obtained for resistive simulations with $\eta=10^{-3}$ and $\eta=10^{-9}$ (not shown here). This implies that the bar instability was unable to generate a sufficiently strong "dynamo" mechanism - i.e., a mechanism that allows for exponential growth - to dominate over the diffusive terms.

\begin{figure}
    \centering
    \includegraphics[width=0.45\textwidth]{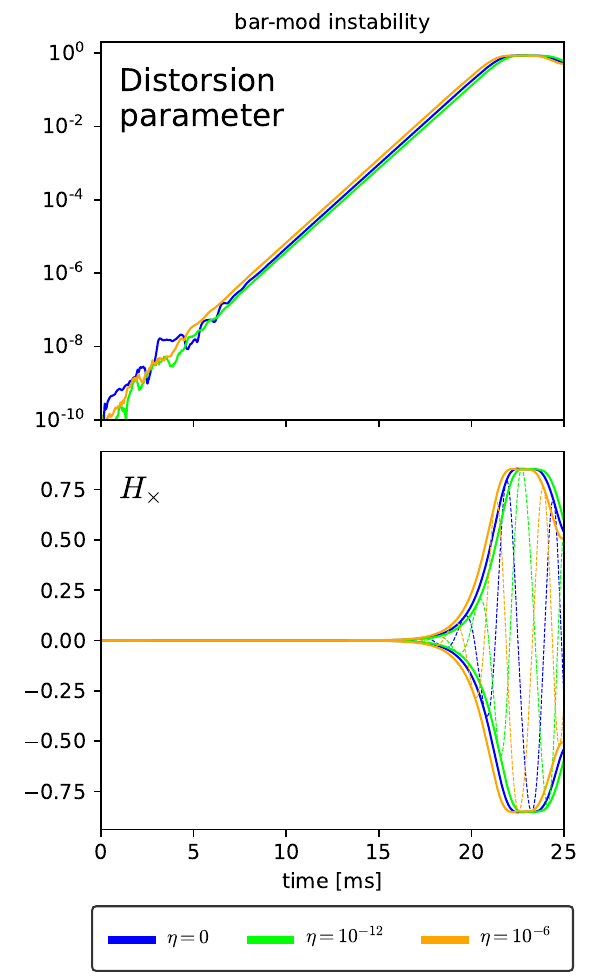}
    \vspace{-2mm}
    \caption{Time evolution of the distortion parameter $H$ (top panel; base-10 log scale) and its $H_{\times}$ component (bottom panel) for the bar-instability simulation in the IMHD and resistive regimes. In the bottom panel, dashed lines represent the $H_{\times}$ component, while the solid line is the distortion factor.}
    \label{fig:u13_imhd_vs_res_quadrupole}.
\end{figure}

Although it significantly affects the evolution of electromagnetic fields, the introduction of physical resistivity does not modify that of the fluid counterpart, as illustrated in Figure \ref{fig:u13_imhd_vs_res_quadrupole}, which shows the time evolution of the distortion parameter $H$, defined as \parcite{Saijo2001,Franci2013}
\begin{equation}
    H=\sqrt{H_+^2+H_{\times}^2}
\end{equation}
with
\begin{equation}
    H_+=\dfrac{I^{xx}-I^{yy}}{I^{xx}+I^{yy}}
\end{equation}
\begin{equation}
    H_{\times}=\dfrac{2I^{xy}}{I^{xx}+I^{yy}}
\end{equation}
where
\begin{equation}
    I^{jk}=\int \dd[3]x \sqrt{\gamma}\,Dx^jx^k
\end{equation}
is the quadrupole moment of the matter distribution. As can be observed, no significant differences are found between the ideal and resistive cases, indicating that the bar instability develops and evolves independently of the electromagnetic fields.

Finally, since these simulations did not require the addition of artificial dissipation, it implies that the physical resistivity is sufficient to suppress the numerical instabilities naturally present in ideal simulations.

%%%%%%%%%%%%%%%%%%%%%%%%%%%%%%%%%%%%%%%%%%%%%%%%%%%%%%%
\section{Conclusions} \label{sec5}
In this work, we presented a new numerical code, named \MIR, developed to solve the equations of general relativistic magnetohydrodynamics (GRMHD) using the 3+1 formalism \parcite{Alcubierre2008,Baumgarte2010,Gourgoulhon2012}. The code operates in Cartesian coordinates and on dynamical backgrounds and has been integrated into the EinsteinToolkit framework \parcite{Loffler2012}, solidifying its applicability in advanced relativistic simulations. A distinctive feature of the \MIR{} code is its ability to solve the GRMHD equations in the isotropic resistive regime, respecting Taub's fundamental inequality \parcite{Taub1948}, thereby ensuring consistency with relativistic kinetic theory.

We also described the numerical implementation of the IMEX SSP2(2,2,2) scheme, necessary for evolving \textit{stiff} equations, as outlined by \cite{Pareschi2005} and \cite{Palenzuela2009}. This scheme has been integrated into the \MoL thorn of the EinsteinToolkit.

The code has been validated through a series of special and general relativity tests. In particular, we demonstrated the three-dimensional evolution of a Tolman-Oppenheimer-Volkoff (TOV) star, both in the presence and absence of a magnetic field, highlighting for the first time the effects of the resistive regime on the electromagnetic field in such configurations. Subsequently, standard tests were conducted on a shock tube and magnetic rotor in two dimensions, in the context of flat spacetime, in both ideal and resistive regimes, and for various electrical resistivity values. The results obtained are in excellent agreement with those produced by other codes. Finally, we performed, for the first time, a three-dimensional simulation of the bar-mode instability in a resistive regime. The results of these simulations demonstrated that implementing physical resistivity not only significantly alters the system's dynamics but also eliminates spurious numerical oscillations, thereby avoiding the need for artificial dissipation, which is typically required in the ideal regime.

For future work, we plan to extend the code by implementing tabulated equations of state to model fluids with more complex properties. Furthermore, we aim to use the \MIR{} code for simulations of binary neutron star mergers in the resistive regime, to study non-ideal effects on the post-merger remnants, as well as the associated electromagnetic and gravitational emissions. 
%%% Lastly, we plan to propose the inclusion of the \MIR{} 
%%% code in future releases of the EinsteinToolkit.

The version of the code used in this study, along with the updated version of the \MoL thorn containing the IMEX SSP2(2,2,2) scheme, is available for free download.

\begin{acknowledgments}

The authors are grateful for the computational resources provided by
CINECA High Performace Comupting (HPC) center, Bologna (Italy) through 
the INFN-CINECA Grant INFN-NEUMAT.

\end{acknowledgments}

%%%%%%%%%%%%%%%%%%%%%%%%%%%%%%%%%%%%%%%%%%%%%%%%%%%%%%%
\section*{Download}
The \MIR{} code can be free downloaded at the following link: \url{https://thegravitationalapple.com/MIR/}. At the same link it is possible to download two other thorns necessary for its operation: a version of \MoL in which the SSP2(2,2,2) scheme has been implemented, and the \texttt{ElectroBase} thorn, in which the matrices for the electric field, the electric charge, the current density and the electrical resistivity have been defined.

%%%%%%%%%%%%%%%%%%%%%%%%%%%%%%%%%%%%%%%%%%%%%%%%%%%%%%%
%%%%%%%%%%%%%%%%%%%%%%%%%%%%%%%%%%%%%%%%%%%%%%%%%%%%%%%
%%%%%%%%%%%%%%%%%%%%%%%%%%%%%%%%%%%%%%%%%%%%%%%%%%%%%%%
\appendix
%%%%%%%%%%%%%%%%%%%%%%%%%%%%%%%%%%%%%%%%%%%%%%%%%%%%%%%
\section{IMEX scheme}\label{app_IMEX}
In this Appendix we report the steps required by the IMEX scheme in the general case, following the prescription presented in \cite{Tomei2019}, itself based on \cite{Bucciantini2013} and \cite{DelZanna2018}. We split the conserved variables \eqref{eq:conserved_variables} as $\UU=\qty{\XX,\YY}$, where $\XX$ is the set of conserved variables with stiff source terms and $\YY$ refers to the remaining ones. In our case $\XX=\sqrt{\gamma}\EE$, since only the source term of the electric field is stiff. In other worlds, $\RR_{\YY}=\zerovec$. It is then convenient to rewrite system \eqref{eq:conservatives_equations} as
\begin{equation}
    \partial_t\XX=\QQ_{\XX}\qty[\UU]+\RR_{\XX}\qty[\UU] \qquad \partial_t\YY=\QQ_{\YY}\qty[\UU]
\end{equation}
Suppose now that in a time interval $\Delta t$ we want to update the conserved variables from $\UU^n$ to $\UU^{n+1}$. The IMEX scheme consist in the following steps:
\begin{itemize}
    \item First, for each step $i=1,2,\ldots,s$ (with $s$ the number of IMEX RK substeps) we have
        \begin{equation}
            \XX_*^{(i)}=\XX^n+\Delta t\sum_{j=1}^{i-1}\at_{ij}\QQ_{\XX}\qty[\UU^{(j)}]+\Delta t\sum_{j=1}^{i-1}a_{ij}\RR_{\XX}\qty[\UU^{(j)}]
        \end{equation}
        and
        \begin{equation}
            \YY_*^{(i)}=\YY^n+\Delta t\sum_{j=1}^{i-1}\at_{ij}\QQ_{\YY}\qty[\UU^{(j)}]
        \end{equation}
        where $\at_{ij}$ and $a_{ij}$ are lower triangular matrices with dimensions $s\times s$.
    \item Second, for $j=i$ variables $\XX_*^{(i)}$ undergo by definition an extra \textit{implicit} evolution with $\at_{ii}=0$ and $a_{ii}\neq 0$:
        \begin{equation}
            \XX^{(i)}=\XX_*^{(i)}+a_{ij}\Delta t\RR_{\XX}\qty[\XX^{(i)},\YY^{(i)}_*] \qquad \YY^{(i)}=\YY^{(i)}_*
        \end{equation}
        Notice that for $i=1$ only the implicit step is needed.
    \item Finally, the conserved variables are updated as
        \begin{equation}
            \UU^{n+1}=\UU^n+\Delta t\sum_{i=1}^s \qty[\bt_i \QQ\qty[\UU^{(i)}]+b_i \RR\qty[\UU^{(i)}]]
        \end{equation}
        where $\bt_i$ and $b_i$ are additional coefficient required by the scheme.
\end{itemize}

%%%%%%%%%%%%%%%%%%%%%%%%%%%%%%%%%%%%%%%%%%%%%%%%%%%%%%%
\section{Derivation of Eqs. \eqref{eq:M_max} and \eqref{eq:S2_max}}\label{app_maxMS2Estar}
\subsection{Maximum for $\boldsymbol{M}$}
From Eq. \eqref{eq:S-definition} we have
\begin{equation}
    S^2=M^2+\norm{\EE\cp\BB}^2+2\Mcon\cdot\qty(\EE\cp\BB)
    \label{eq:app_maxMS2Estar-eq1}
\end{equation}
therefore
\begin{equation}
    S^2\geq M^2+2\Mcon\cdot\qty(\EE\cp\BB)
    \label{eq:app_maxMS2Estar-eq2}
\end{equation}
From Eq. \eqref{eq:E-implicit_step} we obtain
\begin{equation}
\vv\cdot\qty(\EE\cp\BB)=A_0\vv\cdot\qty(\EE_*\cp\BB)+A_1\norm{\vv\cp\BB}^2\geq A_0\vv\cdot\qty(\EE_*\cp\BB)
\end{equation}
Multiplying both sides by $DhW$ we have
\begin{equation}
    \Mcon\cdot\qty(\EE\cp\BB)\geq A_0\Mcon\cdot\qty(\EE_*\cp\BB)
\end{equation}
Since $\Mcon\cdot\qty(\EE_*\cp\BB)\geq -E_*BM$ we obtain
\begin{equation}
    \Mcon\cdot\qty(\EE\cp\BB)\geq -A_0E_*BM
\end{equation}
Furthermore, since $A_0\leq A_{0,\max}$ - which implies $-A_0\geq -A_{0,\max}$ - we have
\begin{equation}
    \Mcon\cdot\qty(\EE\cp\BB)\geq -A_{0,\max}E_*BM
\end{equation}
Inequality \eqref{eq:app_maxMS2Estar-eq2} then becomes
\begin{equation}
    M^2-2A_{0,\max}BE_*M-S^2\leq 0
\end{equation}
Solving for the variable $M$ gives Eq. \eqref{eq:M_max}.

\subsection{Maximum for $\boldsymbol{S^2}$}
In \cite{Etienne2012} is shown that if the dominant energy condition holds - i.e. if $p^2\leq\rho^2\qty(1+\epsilon)^2$ - one has
\begin{equation}
    M^2\leq\en_F^2
    \label{eq:app_maxMS2Estar-eq8}
\end{equation}
From Eq. \eqref{eq:En-definition} we have
\begin{equation}
    \en_F=\en-\Uem\leq\en
\end{equation}
then we can rewrite the inequality \eqref{eq:app_maxMS2Estar-eq8} as
\begin{equation}
    M^2\leq\en^2
\end{equation}
and the Eq. \eqref{eq:app_maxMS2Estar-eq1} becomes
\begin{equation}
    S^2\leq\en^2+\norm{\EE\cp\BB}^2+2\Mcon\cdot\qty(\EE\cp\BB)
    \label{eq:app_maxMS2Estar-eq11}
\end{equation}
Since
\begin{equation}
    \norm{\EE\cp\BB}^2\leq E^2B^2
\end{equation}
and
\begin{equation}
    \Mcon\cdot\qty(\EE\cp\BB)\leq M\norm{\EE\cp\BB}\leq MEB\leq \en EB
\end{equation}
inequality \eqref{eq:app_maxMS2Estar-eq11} becomes
\begin{equation}
    S^2\leq\en^2+E^2B^2+2\en EB
    \label{eq:app_maxMS2Estar-eq14}
\end{equation}
However from the definition of the electromagnetic energy density we have
\begin{equation}
    E^2=2\Uem-B^2\leq 2\Uem
\end{equation}
But, from Eq. \eqref{eq:En-definition},
\begin{equation}
    \Uem=\en-\en_F\leq\en
\end{equation}
then
\begin{equation}
    E^2\leq 2\en
    \label{eq:app_maxMS2Estar-eq17}
\end{equation}
Substituting Eq. \eqref{eq:app_maxMS2Estar-eq17} in Eq. \eqref{eq:app_maxMS2Estar-eq14} we obtain
\begin{equation}
    S^2\leq\en^2+2\en B\sqrt{2\en}+2\en B^2=\qty(\en+B\sqrt{2\en})^2
\end{equation}
that is Eq. \eqref{eq:S2_max}.

%%%%%%%%%%%%%%%%%%%%%%%%%%%%%%%%%%%%%%%%%%%%%%%%%%%%%%%
\section{Equations of State (EoS)}\label{app_eos}
To derive primitive from conservative variables a relation between pressure, mass density and internal energy is needed. This is called \textit{Equation of State} (EoS). In \MIR{} we implemented the following equations:
\begin{itemize}
    \item Polytropic, used only in the atmosphere (see Section \ref{sec:atmosphere}).
    \item Ideal Fluid:
        \begin{equation}
            p=\qty(\Gamma-1)\rho\epsilon
            \label{eq:app_eos-eq1}
        \end{equation}
        where $\Gamma>1$ is a constant.
    \item Taub (e.g. \cite{Mignone2007,Mattia2023}):
        \begin{equation}
            p=\dfrac{\rho\epsilon}{3}\dfrac{\epsilon+2}{\epsilon+1}
            \label{eq:app_eos-eq2}
        \end{equation}
\end{itemize}

During the C2P we need to know the minimum and maximum allowable values for density and internal energy (see Section \ref{sec:C2P}). If a tabulated EoS is used, these values can be obtained from the table.  However, Ideal Fluid and Taub equations theoretically have no upper limit. For this reason, we set $\rho_{\max}=0.0048572$ and $\epsilon_{\max}^{(0)}=51$ as in \cite{Kastaun2021}, while the lower limits are set at atmospheric values.

In \MIR{} we have also implemented the possibility - disabled by default - to enforce the respect of the Taub's fundamental inequality \parcite{Taub1948}
\begin{equation}
    (h-\Theta)(h-4\Theta)\geq 1
    \label{eq:app_eos-eq3}
\end{equation}
- with $\Theta=p/\rho$ the temperature function - which is required to ensure consistency with the kinetic relativistic theory. Eq. \eqref{eq:app_eos-eq2} was derived from Eq. \eqref{eq:app_eos-eq3} by imposing the equal sign, so it satisfies this relation. For the $\Gamma$-law EoS, instead, Taub's inequality implies $\Gamma<5/3$. For $\Gamma\leq 4/3$ the inequality is always fulfilled, otherwise it implies
\begin{equation}
    \epsilon\leq\dfrac{5-3\Gamma}{3\Gamma-4}\equiv\epsilon_{\max}^{(1)} \quad \text{with} \quad \Gamma\in\qty(\dfrac{4}{3},\dfrac{5}{3})
\end{equation}
from which
\begin{equation}
    p\leq\dfrac{\qty(5-3\Gamma)\qty(\Gamma-1)}{3\Gamma-4}\rho
\end{equation}
In this case the maximum value for the internal energy results to be
\begin{equation}
    \epsilon_{\max}=\min\qty{\epsilon_{\max}^{(0)},\epsilon_{\max}^{(1)}}
\end{equation}

%%%%%%%%%%%%%%%%%%%%%%%%%%%%%%%%%%%%%%%%%%%%%%%%%%%%%%%
\section{Divergence of the electric field}\label{app_divergence}
In the 3+1 formalism Eq. \eqref{eq:Maxwell-Fmunu-covariant} becomes
\begin{equation}
    \nabla_{\mu}F^{\mu\nu}=-qn^{\nu}-J^{\nu}
\end{equation}
The parallel (time) projection gives
\begin{equation}
    \gamma^{-1/2}\partial_{\mu}\qty(\sqrt{\gamma}\,E^{\mu})=q
\end{equation}
that is
\begin{equation}
    \partial_t\qty(\sqrt{\gamma}\,E^0)+\partial_i\qty(\sqrt{\gamma}\,E^i)= \sqrt{\gamma}\,q
\end{equation}
Since $E^0=0$ one has
\begin{equation}
    \partial_t0+\partial_i\qty(\sqrt{\gamma}\,E^i)=\sqrt{\gamma}\,q
    \label{eq:app_divergence-eq4}
\end{equation}
Since $\partial_t0=0$ we have Eq. \eqref{eq:Maxwell1}. However, Eq. \eqref{eq:app_divergence-eq4} can be read as the equation for the evolution of the conserved variable $\UU=0$. We can then compute the spatial derivative $\partial_i$ using Eq. \eqref{eq:HLLE} with $\UU^i_{R,j}=\UU^i_{L,j}=0$.

%%%%%%%%%%%%%%%%%%%%%%%%%%%%%%%%%%%%%%%%%%%%%%%%%%%%%%%
%%%%%%%%%%%%%%%%%%%%%%%%%%%%%%%%%%%%%%%%%%%%%%%%%%%%%%%
%%%%%%%%%%%%%%%%%%%%%%%%%%%%%%%%%%%%%%%%%%%%%%%%%%%%%%%
%% \printbibliography

\bibliographystyle{apsrev4-1}

%%\bibliography{bibliografia}
%%\end{document}

%https://arxiv.org/pdf/1105.1895.pdf
%https://journals.aps.org/prd/pdf/10.1103/PhysRevD.103.023018

%https://iopscience.iop.org/article/10.3847/1538-4365/ab3922


\end{document}